\documentclass[epj]{svjour}

\pdfoutput=1

\RequirePackage[numbers,sort&compress]{natbib}
\usepackage[T1]{fontenc} 
\usepackage{amsmath}
\usepackage{amssymb}
\usepackage[usenames,dvipsnames]{color}
\include{jdefs}
\usepackage{multirow}
\usepackage{graphicx}
\usepackage{adjustbox}
\usepackage{booktabs}
\usepackage{appendix}
\usepackage{slashed}
\usepackage{url}
\usepackage{xspace}
\usepackage{braket}
\usepackage{tabularx}
\usepackage{enumerate}
\usepackage{widetext}

\definecolor{darkblue}{rgb}{0,0,0.5}
\definecolor{darkgreen}{rgb}{0.1,0,0.3}
\definecolor{darkred}{rgb}{0.6,0,0}
\RequirePackage[bookmarks=false, bookmarksnumbered=true, colorlinks=true, urlcolor=darkblue, citecolor=darkgreen, linkcolor=darkred, breaklinks=true]{hyperref}
\usepackage[normalem]{ulem} 
\usepackage[utf8]{inputenc}
\usepackage{bm}
\usepackage{siunitx}
\usepackage{commath}
\usepackage{mathtools}
\usepackage{cuted}
\usepackage[capitalise]{cleveref}
\usepackage{widetext}

\newcommand{\Umt}{$U(1)_{L_\mu-L_\tau}$\xspace}
\newcommand{\Um}{$U(1)_{L_\mu}$\xspace}
\newcommand{\gmu}{$(g-2)_\mu$\xspace}
\newcommand{\NAmu}{NA64$\mu$\xspace}
\newcommand{\cevns}{CE$\nu$NS\xspace}

\newcommand{\knr}{\mathrm{keV}_\mathrm{nr}\xspace}
\newcommand{\kee}{\mathrm{keV}_\mathrm{ee}\xspace}

\newcommand{\ee}{\textsubscript{ee}}
\newcommand{\nr}{\textsubscript{nr}}

\begin{document}



\title{
Confirming $U(1)_{L_\mu-L_{\tau}}$ as a solution for $(g-2)_\mu$ with neutrinos
}
\date{Received: date / Revised version: date}

 \author{D.W.P. Amaral\inst{1}, D.G. Cerde\~no\inst{2,3}, A. Cheek\inst{4}, and P. Foldenauer\inst{1}}
 \institute{Institute for Particle Physics Phenomenology, Durham University, Durham DH1 3LE, United Kingdom \and Instituto de F\' isica Te\'orica, Universidad Aut\'onoma de Madrid, 28049 Madrid, Spain \and Departamento de F\' isica Te\'orica, Universidad Aut\'onoma de Madrid, 28049 Madrid, Spain \and Centre for Cosmology, Particle Physics and Phenomenology (CP3), Universit\'e Catholique de Louvain, Chemin du Cyclotron 2, B-1348 Louvain-la-Neuve, Belgium}

\abstract{
The recent measurement of the muon anomalous magnetic moment by the Fermilab E989 experiment, when combined with the previous result at BNL, has confirmed the tension with the SM prediction at $4.2\,\sigma$~CL, strengthening the motivation for new physics in the leptonic sector. Among the different particle physics models that could account for such an excess, a gauged $U(1)_{L_\mu-L_{\tau}}$ stands out for its simplicity. In this article, we explore how the combination of data from  different future probes can help identify the nature of the new physics behind the muon anomalous magnetic moment. In particular, we contrast $U(1)_{L_\mu-L_{\tau}}$ with an effective $U(1)_{L_\mu}$-type model. We first show that muon fixed target experiments (such as NA64$\mu$) will be able to measure the coupling of the hidden photon to the muon sector in the region compatible with $(g-2)_\mu$, and will have some sensitivity to the hidden photon's mass. We then study how experiments looking for coherent elastic neutrino-nucleus scattering (CE$\nu$NS) at spallation sources will provide crucial additional information on the kinetic mixing of the hidden photon. When combined with NA64$\mu$ results, the exclusion limits (or reconstructed regions) of future CE$\nu$NS detectors will also allow for a better measurement of the mediator mass.
Finally, the observation of nuclear recoils from solar neutrinos in dark matter direct detection experiments will provide unique information about the coupling of the hidden photon to the tau sector. The signal expected for $U(1)_{L_\mu-L_{\tau}}$ is larger than for $U(1)_{L_\mu}$ with the same kinetic mixing, and future multi-ton liquid xenon proposals (such as DARWIN) have the potential to confirm the former over the latter. We determine the necessary exposure and energy threshold for a potential $5\,\sigma$ discovery of a $U(1)_{L_\mu-L_{\tau}}$ boson, and we conclude that the future DARWIN observatory will be able to carry out this measurement if the experimental threshold is lowered to $1\,\knr$.
} 

\authorrunning{D.W.P. Amaral et al.}
\titlerunning{Confirming $U(1)_{L_\mu-L_{\tau}}$ as a solution for $(g-2)_\mu$ with neutrinos}
\maketitle

\section{Introduction}
\label{sec:introduction}

The recent determination of the muon anomalous magnetic moment, $a_\mu = (g-2)_\mu/2$, by the E989 experiment at Fermilab~\cite{Abi:2021gix} shows a discrepancy with respect to the theoretical prediction of the Standard Model (SM)~\cite{Aoyama:2012wk,Aoyama:2019ryr,Czarnecki:2002nt,Gnendiger:2013pva,Davier:2017zfy,Keshavarzi:2018mgv,Colangelo:2018mtw,Hoferichter:2019gzf,Davier:2019can,Keshavarzi:2019abf,Kurz:2014wya,Melnikov:2003xd,Masjuan:2017tvw,Colangelo:2017fiz,Hoferichter:2018kwz,Gerardin:2019vio,Bijnens:2019ghy,Colangelo:2019uex,Blum:2019ugy,Colangelo:2014qya}\footnote{More details on the individual contributions entering the SM calculations can be found in Ref.~\cite{Aoyama:2020ynm}. Furthermore, it should be noted that a recent lattice calculation of the leading order hadronic vacuum polarisation~\cite{Borsanyi:2020mff} significantly reduces this difference; however, this worsens fits to other precision EW observables~\cite{Passera:2008jk,Crivellin:2020zul}.},
\begin{align}
     a_\mu^\mathrm{FNAL} &=  116\ 592\ 040(54) \times 10^{-11}\,, \nonumber \\[.2cm]
     a_\mu^\mathrm{SM} &=  116\ 591\ 810(43) \times 10^{-11}\,.
\end{align}
When combined with the previous Brookhaven determination~\cite{Bennett:2002jb,Bennett:2004pv,Bennett:2006fi} of
\begin{equation}
    a_\mu^\mathrm{BNL} =  116\ 592\ 089(63) \times 10^{-11}\,,
\end{equation}
this leads to a $4.2\,\sigma$ observed excess of 
\begin{equation}\label{eq:g2excess}
    \Delta a_\mu = 251(59)\times 10^{-11} \,.
\end{equation}

This exciting result represents a substantial improvement with respect to the previous measurement at BNL~\cite{Bennett:2002jb,Bennett:2004pv,Bennett:2006fi}, as the experimental uncertainty has been reduced. Since the central value has shifted towards the SM value, the resulting discrepancy falls short of confirming this as evidence for new (BSM) physics, although it certainly leaves the door open for this exciting possibility. While more data from E989 is needed to ultimately judge whether this excess is due to a statistical fluctuation  or new physics, complementary search strategies for the possible origin of this excess should be pursued.

During the past two decades, different realisations of particle physics models have been proposed to address this tension, which in general consider extensions in the leptonic sector. Among these, the gauged \Umt stands out for its simplicity, as it is anomaly free without the addition of any extra new fermion fields~\cite{He:1990pn,He:1991qd,Ma:2001md}. Various realisations of this model have been studied, including also additional neutrino~\cite{Ma:2001md,Heeck:2011wj,Covi:2020pch} and dark  matter (DM) fields~\cite{Foldenauer:2018zrz,Okada:2019sbb,Borah:2020jzi,Kamada:2020buc}. Recently, it has also been noted that a simultaneous explanation of the Hubble tension can be achieved in these models \cite{Escudero:2019gzq,Araki:2021xdk}. The purpose of this work is to identify a set of complementary experiments suited for an independent confirmation of \Umt as a solution to the $(g-2)_\mu$ anomaly.

The fundamental difficulty in inferring the underlying model of the \gmu excess is that \gmu is only sensitive to the muon coupling of any new hypothetical mediator. Hence, in order to explain the \gmu excess alone, a simplified mediator model suffices in which the new scalar or vector mediator couples to muons only. Such a new muon-philic singlet mediator will be directly accessible at muon beam experiments. If  indeed the only coupling of the  new mediator to the SM is to muons alone, it can for example be tested in high-energy muonic Bhabha scattering or $\mu\mu\to h \gamma$ at  muon colliders~\cite{Buttazzo:2020eyl,Yin:2020afe,Capdevilla:2021rwo}. However, in order to explain the positive shift in \gmu with respect to the SM, the new interaction must be (mostly) of scalar- or vector-type rather than pseudo-scalar or axial-vector~\cite{Freitas:2014pua,Heeck:2016xkh}. This necessarily implies that the new mediator has to couple to both left- and right-handed components of the muon with approximately equal strength. Since the fundamental building blocks of the SM in terms of matter fields are the irreducible representations of $SU(3)_{C}\times SU(2)_{L}\times U(1)_{Y}$, gauge invariance requires the new mediator at the fundamental level to couple to the left-handed $SU(2)_{L}$ lepton doublet $L$ rather than just the left-handed component of the muon $\mu_L$.

In the case of a new vector mediator, the only gauge-invariant, renormalisable interaction terms required to solve the \gmu anomaly are given by
\begin{equation} \label{eq:muint}
    \mathcal{L}_\mathrm{int} \supset - g_{x}\, \bar L_2 \gamma^\alpha L_2 \, X_\alpha - g_{x}\, \bar \mu_R \gamma^\alpha \mu_R  \, X_\alpha \,,
\end{equation}
where $L^T_2= (\nu_{\mu L}, \mu_L)$ and $g_x$ denotes the coupling of the new vector boson $X_\alpha$. The Lagrangian in~\cref{eq:muint} necessarily implies an equal-strength interaction of the new vector boson with the muon-neutrino as with the muon. In this paper, we study the potential of such neutrino interactions for unveiling the underlying physics behind the \gmu excess, if indeed it is due to a new vector mediator.

The Lagrangian in~\cref{eq:muint} could correspond to the interactions of a $U(1)_{L_\mu}$ gauge boson with the SM. However, such a $U(1)_{L_\mu}$ is anomalous with only the SM field content and new fields need to be added to cancel the anomalies.\footnote{Such anomalous leptophilic vector models of gauged lepton number can play an important role in baryogenesis~\cite{Carena:2018cjh,Carena:2019xrr}.} Sensible UV completions of the \Um model quickly run into the highly restrictive flavour-changing constraints \cite{Carena:2019xrr}.\footnote{We want to thank Yue Zhang for pointing this out to us.} On the other hand, the \Umt model is anomaly-free with only the SM field content and is therefore theoretically very appealing. In this work, we are interested in experimentally confirming \Umt by measuring its three characteristic properties. 
\begin{enumerate}
    \item[\textbf{P1.}] A vector-like coupling to the second generation of leptons.
    \item[\textbf{P2.}] A specific value for the kinetic mixing with the SM photon, namely $\epsilon \sim  g_x/70$.
    \item[\textbf{P3.}] A vector-like coupling to the third generation of leptons, which is equal and opposite to that of the second.
\end{enumerate}

As explained above, a vector solution to the \gmu excess only needs to satisfy property P1, so in order to confirm that \Umt is indeed responsible for the \gmu anomaly, we must verify properties P2 and P3. Since a generic \Um-type mediator satisfies property P1, has the freedom to satisfy property P2, but does not satisfy property P3, it is a good model to contrast with \Umt due to its similar, but not identical, physical predictions.

If the new vector boson exists and is light, in particular lighter than the dimuon threshold $M_X < 2 m_\mu$, the only way to directly search for such a boson is via its invisible decay into neutrinos. This can for example be done by a missing momentum search in muon fixed target experiments like M$^3$~\cite{Kahn:2018cqs} or \NAmu~\cite{Gninenko:2014pea,Gninenko:2018tlp}, where the muon-philic vector is produced via Bremsstrahlung in muon-nucleus collisions.  Additionally, it can also be looked for in the low-energy $e^+e^-$ collider experiment BESIII \cite{Cvetic:2020vkk}. If on the contrary the new boson is heavy, $M_X>2 m_\mu$ one could directly search for it in the visible decay into a pair of muons at \NAmu~\cite{Gninenko:2018tlp} as well as at M$^3$~\cite{Kahn:2018cqs}. Alternatively, the new boson can be produced in kaon decays $K\to\mu\nu X$ and searched for in a missing energy signature at NA62~\cite{Krnjaic:2019rsv}. Hence, these searches would be prime candidates to confirm that the \gmu excess is indeed due to a new muon-philic vector mediator, thus testing P1. However, these  missing energy searches are only sensitive to the production cross section of the muon-philic vector boson $\sigma_X$ and its total invisible branching ratio BR$_{X\to\mathrm{inv}}$. Therefore, these searches cannot distinguish a muon-coupled boson with decays into light dark sector states from e.g.~a \Um or \Umt boson with decays into neutrinos.

In order to test P2, we need to get a handle on the kinetic mixing parameter, $\epsilon$. To this end, we consider experiments measuring coherent elastic neutrino-nucleus scattering (\cevns) at spallation sources. This also allows one to distinguish between neutrino-coupled and DM-coupled mediators because the former would induce corrections to the \cevns rate, while the latter would not. \cevns was first detected at the COHERENT experiment, which employed a CsI[Na] target~\cite{Akimov:2015nza,Akimov:2017ade}. Recently, this observation has been replicated using argon nuclei \cite{Akimov:2020pdx}. Since the results are compatible with the SM theoretical prediction, they can be used to constrain new physics in the neutrino sector. Data from the CsI run was used to derive limits for the \Umt model~\cite{Abdullah:2018ykz} and in Ref.~\cite{Amaral:2020tga} we computed the limits from CENNS-10 LAr results. The combined bounds from CsI and LAr have been presented in Ref.~\cite{Banerjee:2021laz}. The current exclusion line does not probe the \gmu favoured region for \Umt.

In the future, there are plans to build more sensitive multi-ton detectors to further test \cevns and probe neutrino non-standard interactions \cite{Banerjee:2021laz,Miranda:2020syh,Shoemaker:2021hvm}. These include the European Spallation \sloppy Source (ESS) \cite{Baxter:2019mcx}, the Coherent CAPTAIN-Mills (CCM) at Los Alamos National Laboratory \cite{ccm}, and Oak Ridge National Laboratory \cite{Akimov:2019xdj}. In this work, we investigate the sensitivity of all these experiments to \Umt and find that their predicted reach will allow them to test some of the areas of its parameter space which are consistent with the recent \gmu measurement. In the case of detection, these experiments combined with muon beam experiments provide a powerful strategy for constraining the value of $\epsilon$.

It should be noted that all above detectors are only able to probe the second generation lepton couplings, thereby not directly testing P3. In order to confirm a \Umt boson as a solution for \gmu, one therefore needs sensitivity to the third generation lepton couplings. This can be achieved by considering solar neutrino scattering, as the solar neutrino flux has a $\nu_\tau$ component due to neutrino oscillations. In this context, direct detection experiments are becoming increasingly sensitive to rare events, and they will soon have the potential to probe new physics in solar neutrino scattering. Although their main purpose is to observe dark matter particles, the increased payload, together with the improved cleanliness and analysis techniques, will turn the new generation detectors into multi-purpose instruments. In particular, it is expected that they will be able to observe neutrinos through either their scattering off electrons or through \cevns. This is, in fact, often regarded as a limitation for DM searches, since \cevns events will constitute a new background which is difficult to disentangle from potential DM events.

In Ref.~\cite{Amaral:2020tga}, we pointed out that direct detection, and in particular experiments based on liquid xenon, could be crucial to probe the \Umt solution to the muon anomalous magnetic moment. The new results of the E989 experiment confirm this, since the $2\,\sigma$ region in the $(g_{\mu\tau}, M_{A^\prime})$ parameter space lies partly within the expected sensitivity of future detectors such as DARWIN. In this work, we study how well a potential signal from \Umt in agreement with the latest \gmu measurement could be reconstructed in multi-ton liquid xenon experiments. We find that, since the contribution to nuclear recoils from \Umt and \Um differs, direct detection experiments will be crucial to distinguish between these models when combined with results from muon beam and \cevns experiments.

\bigskip

This article is organised as follows. In \cref{sec:models}, we introduce two extensions of the SM with a muon-philic vector mediator, namely a minimal gauged \Umt and a simplified \Um for comparison. We compute the region of the parameter space compatible with the new \gmu measurement and define a set of benchmark points for analysis. In \cref{sec:beam_dump}, we study how this region can be probed at muon beam experiments, specifically at \NAmu, arguing that these experiments will be able to constrain the mass of the mediator and the couplings to the muon sector. Next, in \cref{sec:cevns}, we turn our attention to experiments looking for \cevns in spallation sources, reviewing the current experimental situation and the forecast for future detectors, finding that these experiments can constrain the kinetic mixing $\epsilon$ when combined with muon beam experiments. Finally, in \cref{sec:direct}, we include direct detection probes and study the performance of upcoming and future xenon-based detectors, focusing on the observation of nuclear recoils induced by the \cevns of solar neutrinos. We show that, when combined with the previous experimental analysis, direct detection experiments could confirm a signal from the \Umt model for mediator masses below $\sim 50$~MeV due to their sensitivity to the tau neutrino coupling. Finally, we present our conclusions in \cref{sec:conclusions}.

\section{Muon-philic vector mediators and \gmu}
\label{sec:models}

\begin{figure*}[!t]
    \begin{center}
    \includegraphics[width=0.9\textwidth]{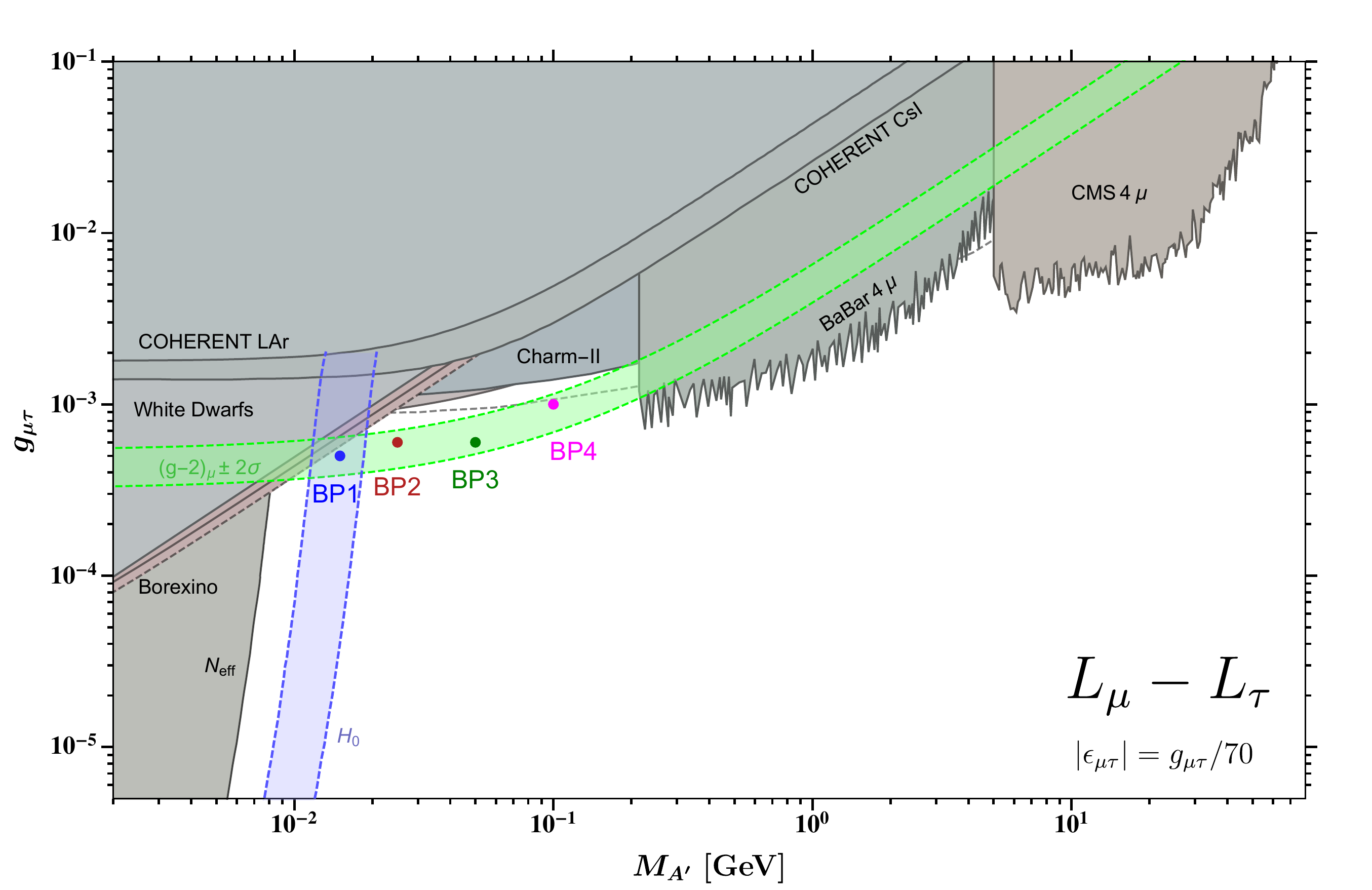}
    \end{center}
    \caption{\label{fig:lmlt_g2} Current constraints on the parameter space of the gauge boson of a minimal \Umt gauge group from $N_\mathrm{eff}$~\cite{Escudero:2019gzq}, Borexino~\cite{Amaral:2020tga}, white dwarf cooling~\cite{Bauer:2018onh}, COHERENT CsI~\cite{Abdullah:2018ykz} and LAr~\cite{Amaral:2020tga}, neutrino tridents (Charm-II)~\cite{Altmannshofer:2014pba,Geiregat:1990gz} and BaBar~\cite{TheBABAR:2016rlg} and CMS~\cite{Sirunyan:2018nnz} 4$\mu$ searches in grey. We show the  neutrino trident constraint from CCFR as a grey dashed line, since some backgrounds have not been properly taken into account~\cite{Krnjaic:2019rsv}. The corresponding plot for a simplified \Um with the same kinetic mixing looks exactly the same with the exception of the BaBar and CMS 4$\mu$ limits, which are slightly shifted towards smaller couplings by a factor $\sqrt{\mathrm{BR}_{A'\to\mu\mu}^{\mu-\tau}/\mathrm{BR}_{A'\to\mu\mu}^{\mu}}\approx$ 0.87 (0.71) below (above) the ditau threshold due to the increased branching ratio of the $L_\mu$ boson to muons. The green band shows the region favoured at $2\sigma$ by the recent confirmation of the $(g-2)_\mu$ excess by the E989 experiment~\cite{Abi:2021gix}. The blue band shows the region favoured by $H_0$~\cite{Escudero:2019gzq}. For illustrative purposes we also show the four benchmark points BP1 - BP4 used for analysis in this work.
}
\end{figure*}

The focus of this paper is to study the physical implications of an explanation of the \gmu excess in terms of a \Umt gauge boson. For this purpose we consider an extension of the SM by a minimal \Umt gauge model and an effective \Um-type model. Throughout this work, we not only test an experiment's ability to reconstruct a signal from a \Umt gauge model, but also test its power to differentiate between the predictions of a \Umt and those of a \Um.

Quite generically, the Lagrangian of a $U(1)_X$ extensions of the SM can be written in the form
\begin{align}\label{eq:u1lag}
    \mathcal{L} = \mathcal{L}_\mathrm{SM} &- \frac{1}{4} X_{\alpha\beta} X^{\alpha\beta} - \frac{\epsilon_Y}{2}  B_{\alpha\beta} X^{\alpha\beta} - g_x \, j^X_\alpha \, X^\alpha \nonumber \\
    &- \frac{M_{X}^2}{2} X_\alpha X^\alpha \, ,
\end{align}
where $B_{\alpha\beta}$ and $X_{\alpha\beta}$ denote the field strength tensors of the SM hypercharge boson $B_\alpha$ and new $U(1)_X$ gauge boson $X_\alpha$, $\epsilon_Y$ is the kinetic mixing parameter, $M_{X}$ is the mass of the new boson, and $g_x$ and $j^X_\alpha$ are the $U(1)_X$ coupling constant and gauge current, respectively. For the two types of models we are considering, the currents are given by 
\begin{align}
    j^{\mu-\tau}_\alpha & =  \bar L_2 \gamma_\alpha L_2 
         + \bar \mu_R \gamma_\alpha \mu_R 
         - \bar L_3 \gamma_\alpha L_3 -\bar\tau_R \gamma_\alpha \tau_R\,, \\
    j^{\mu}_\alpha &= \bar L_2 \gamma_\alpha L_2 
         + \bar \mu_R \gamma_\alpha \mu_R 
         +\sum_{\psi} Q_\psi \bar \psi \gamma_\alpha \psi  \,,
\end{align}
where $L_i^T= (\nu_i, e_i)$ denotes the lepton doublet of the $i$th generation and $\psi$ are  new heavy fields needed to cancel the anomalies in the case of \Um. 

It is worth pointing out that effective \Um models as an explanation of \gmu as considered here can be subject to very strong constraints from flavour-changing processes like $K\to \pi X$ and $B\to K X$~\cite{Carena:2019xrr}. These constraints are due to enhanced production of the longitudinal mode of the  associated gauge boson coupled to an anomalous current~\cite{Dror:2017nsg}. In the low-energy EFT, the heavy fields $\psi$ (which carry electroweak charge) can be integrated out to produce Wess-Zumino (WZ) terms coupling the new $X$-boson to the weak bosons, leading to flavour-changing penguin diagrams. The coefficients for the WZ terms are dictated by the underlying UV completion of \Um and are, in general, non-vanishing. 
 
Since we are using the \Um model as a mere foil for the \Umt model, we are not interested in its possible UV completions, and we rather view it as an effective model for a muon-coupled vector mediator. If indeed experimental results begin to favour \Um, this could indicate some finely tuned cancellation of flavour-changing currents. Alternatively, there could be a scenario where the strong flavour constraints are absent. This could be realised with UV-completing heavy fields $\psi$ that are SM-chiral with masses that break the electroweak symmetry~\cite{Dror:2017nsg}. The new fields $\psi$ would enter the spectrum at a new physics scale $M_\mathrm{NP}$ somewhere between the electroweak scale $v=246$ GeV and a possible GUT scale $f_\mathrm{GUT}\approx 10^{16}$ GeV.\footnote{In particular, the mixed $[SU(2)]^2[U(1)]$ anomaly does not cancel in \Um with the SM content alone and therefore one has to add new fields transforming non-trivially under $SU(2)_L$. In this context, an interesting direction could be to revive fourth generation models with an extended Higgs sector~\cite{Lenz:2013iha,Das:2017mnu}, where the fourth generation leptons are charged under a $U(1)_{L_\mu -L_4}$ symmetry.} An extension of the SM by a gauge \Umt is already anomaly-free with only the field content of the SM~\cite{He:1990pn,He:1991qd,Ma:2001md}, but it can even be extended by three right-handed neutrinos to a minimal physically viable model for neutrino masses with correct prediction of the CKM and PMNS matrices~\cite{Heeck:2011wj,Araki:2012ip,Kownacki:2016pmx,Asai:2017ryy,Bauer:2020itv}.

In order to make contact with experiments, we have to canonically normalise the fields in~\cref{eq:u1lag} and change to the mass-basis of the fields. This can be done by diagonalising the kinetic terms of the gauge bosons via a field redefinition and two consecutive rotations outlined in Ref.~\cite{Bauer:2018onh}, leading to the interaction terms in the form
\begin{equation}
    \mathcal{L}_\mathrm{int} = - (e\, j^\mathrm{em}_\alpha, g_Z \, j^Z_\alpha, g_x \, j^X_\alpha) \ K \ \left(\begin{matrix}
    A^\alpha \\
    Z^\alpha \\
    A'^\alpha
    \end{matrix}\right) \,,
\end{equation}
where $e$ denotes the electromagnetic coupling constant and $g_Z=e/(\sin\theta_W \cos\theta_W)$ with the weak mixing angle $\theta_W$. Here $j^\mathrm{em}_\alpha$ and $j^Z_\alpha$ are the electromagnetic and weak neutral current, and $A_\alpha$, $Z_\alpha$ and $A'_\alpha$ are the mass eigenstates of the SM photon, weak neutral boson and the new gauge boson, which we will refer to as {\it hidden photon}. The coupling matrix $K$ is approximately given by~\cite{Bauer:2018onh}
\begin{equation}
    K \simeq \left(\begin{matrix}
    1 & 0 & -\epsilon_x \\
    0 & 1 & 0 \\
    0 & \epsilon_x \tan\theta_W  & 1
    \end{matrix}\right)\,,
\end{equation}
with the physical kinetic mixing parameter $\epsilon_x=\epsilon_Y \,\cos\theta_W$.

In this article, we will assume that there is no kinetic mixing at tree level, either because a mixing term is forbidden by the underlying UV symmetries or because the tree-level mixing is much smaller than the loop-induced one. At low energies $q\ll v_\mathrm{EW}$ the mixing of the new $X$-boson in~\cref{eq:u1lag} is to good approximation only with the SM photon $A$. In this limit of a low-energy effective theory, the kinetic mixing parameter induced at one-loop can be expressed as
 \begin{align}
     \epsilon_x(q) = - \frac{e g_x}{2 \pi^2} &\int^1_0 \dif x\, \left\{ x (1-x) \sum_f Q_f Q_f^x \nonumber \right. \\ &\times \left.\log\left(\frac{\mu^2}{m_f^2-x(1-x)q^2}\right)\right\}\,,
 \end{align}
where  $f$ are the fermions running in loop, $Q_f$, $Q_f^x$ and $m_f$ denote their electromagnetic,  $U(1)_X$ charges and mass, respectively, and $\mu$ is a renormalisation scale. For our models, the kinetic mixing parameters  can then be approximated at energies below the muon mass, $q^2\ll m_\mu^2$, by
\begin{align}
   \epsilon_{\mu\tau} &\approx \frac{e \, g_{\mu \tau}}{6\pi^2}\log\left(\frac{m_\mu}{m_\tau} \right) \approx - \frac{g_{\mu\tau}}{70}\,, 
   \label{eq:epsilonmutau}\\
   \epsilon_{\mu} &\approx \frac{e \, g_{\mu}}{6\pi^2}\log\left(\frac{m_\mu}{M_{\rm NP}} \right)\approx - Log \ g_\mu\,, 
   \label{eq:epsilonmu}
\end{align}
where the numerical prefactor $Log\approx 10^{-2}-10^{-1}$ for a new physics scale for the fields $\psi$ of $M_{\rm NP} =10^2-10^{16}$ GeV. So, even if we are agnostic about the true mass scale of the new fields $\psi$, we can get a rough estimate of the loop-induced kinetic mixing in the case of a UV completion of \Um.

\subsection{Muon anomalous magnetic moment}
\label{sec:g-2}

\begin{figure}[!t]
    \begin{center}
    \includegraphics[width=.4\textwidth]{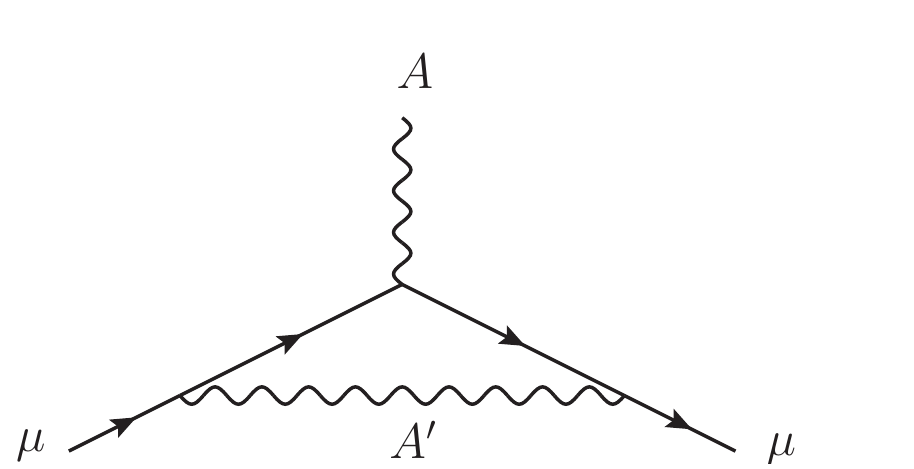}
    \end{center}
    \caption{\label{fig:g2} Contribution of the \Umt gauge boson to the anomalous magnetic moment of the muon.}
\end{figure}

The muon-philic gauge bosons of \Um and \Umt lead to a shift in the anomalous magnetic of the muon\footnote{Note that a similar contribution exists for the anomalous magnetic moment of the electron $a_e$, which is, however, doubly suppressed by kinetic mixing. Since we are only considering loop-induced kinetic mixing, this corresponds effectively to a three-loop process. In this context, it is interesting to note that the latest experimental determination of $(g-2)_e$ shows a mild deficit compared to the SM prediction~\cite{Parker:2018vye}. Due to the positive sign of the shift from vectorial couplings, the corresponding contribution via a \Um or \Umt boson slightly worsens this tension.}, $a_\mu = (g-2)_\mu/2$, induced by the loop diagram shown in~\cref{fig:g2}. Quite universally, the  shift $\Delta a_\mu$ induced by a neutral gauge boson with vectorial couplings to muons (as in \Um and \Umt) can be expressed in the compact form~\cite{Lynch:2001zs,Pospelov:2008zw}
\begin{equation}\label{eq:g2mu}
    \Delta a_\mu =Q_\mu^{x^2} \ \frac{\alpha_x}{\pi} \int_0^1 \dif u \ \frac{u^2(1-u)}{u^2+\frac{(1-u)}{x_\mu^2}}\,,
\end{equation}
where $\alpha_x=g_{x}^2/4\pi$, $x_\mu=m_\mu/M_{A'}$ and $Q_\mu^x$ denotes the charge of the muon under the new gauge group.

Since in both \Um and \Umt the charge of the muon is fixed at $Q_\mu^x = 1$, the observed excess in~\cref{eq:g2excess} translates directly to a band of preferred coupling $g_x$ and mass $M_{A'}$ values in the 2D parameter as shown in~\cref{fig:lmlt_g2}. This defines a clear target parameter space for where to look for this kind of new physics. In this paper, we want to explore the next steps necessary to test the hypothesis that the observed \gmu excess is indeed due to a \Umt hidden photon by exploring this target region with special emphasis on the additional insights gained by exploiting the neutrino interactions of these bosons.

\subsection{Benchmark points and analysis strategy}
\label{sec:bps}

\begin{table}[h]
    \centering
    \begin{tabular}{c|c|c}
    \hline\hline
    & $M_{A'}$& $g_{\mu\tau}$\\
    \hline
        BP1 & 15~MeV & $5\times 10^{-4}$ \\[1ex]
        BP2 & 25~MeV & $6\times 10^{-4}$ \\[1ex]
        BP3 & 50~MeV & $6\times 10^{-4}$ \\[1ex]
        BP4 & 100~MeV & $1\times 10^{-3}$ \\[1ex]
        \hline\hline
    \end{tabular}
    \caption{Benchmark points in the $(g_{\mu\tau}, M_{A'})$ parameter space of a \Umt boson favoured by \gmu. For \Umt we use the value of the loop-induced kinetic mixing, $\epsilon_{\mu\tau} = -g_{\mu\tau}/70$. }
    \label{tab:bps}
\end{table}

In this work, we want to explore the sensitivities of various different experiments like muon beam, coherent elastic neutrino nucleus scattering (\cevns) and DM direct detection experiments to a solution of \gmu in terms of a muon-philic hidden photon $A'$. For this purpose, we define the four benchmark points (BP) of~\cref{tab:bps} in the region of parameter space favoured by \gmu in a \Umt model. The benchmark points are also illustrated in~\cref{fig:lmlt_g2}.

Throughout this paper, we study the sensitivities of the different experiments to a hypothetical signal coming from a \Umt boson with coupling and mass as defined by the benchmark points. For all experiments, we therefore generate mock data sets for the four different \Umt benchmark points and try to reconstruct the mass $M_{A'}$ and coupling $g_x$ under both the hypothesis that the signal is due to a \Umt or due to a \Um hidden photon. In the remainder of this paper, we will refer collectively to the coupling and mixing of \Um and \Umt with $g_x$ and $\epsilon_x$, and label them by an index $\mu$ or $\mu\tau$ when we need to distinguish between these groups.

\section{Muon beam experiments}
\label{sec:beam_dump}

With the recent measurement of the E989 experiment confirming the \gmu excess,  independent complementary measurements will be of paramount importance for pinning down its true nature. If the excess is indeed due to a new light muon-philic vector boson, a missing energy search at muon beam dump experiments will be crucial in confirming this. Both the planned CERN \NAmu~\cite{Gninenko:2640930,Gninenko:2653581} and Fermilab $M^3$~\cite{Kahn:2018cqs} experiments are prime candidates to conduct such a search. The \NAmu experiment is planned to perform a first pilot run in 2021~\cite{Gninenko:2020hbd} and conduct its full Phase-I run in 2021-2023~\cite{Gninenko:2653581} with a total of $10^{11}$ muons on target (MOT). This will therefore be the first experiment to be sensitive to the full parameter space of a muon-philic boson allowed by \gmu. We therefore constrain our analysis of missing energy searches at muon beam experiments to the case of \NAmu, but note that our results should be easily generalised to the case of the Fermilab $M^3$ experiment.

\begin{figure}[!h]
    \begin{center}
    \includegraphics[width=.40\textwidth]{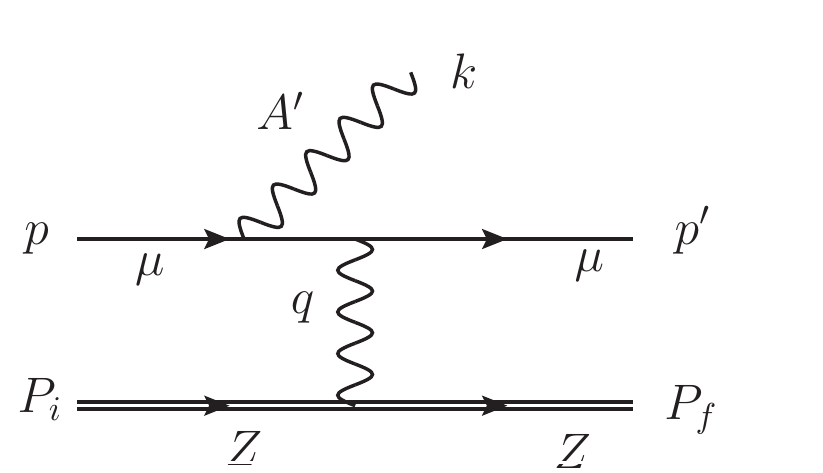}
    \includegraphics[width=.40\textwidth]{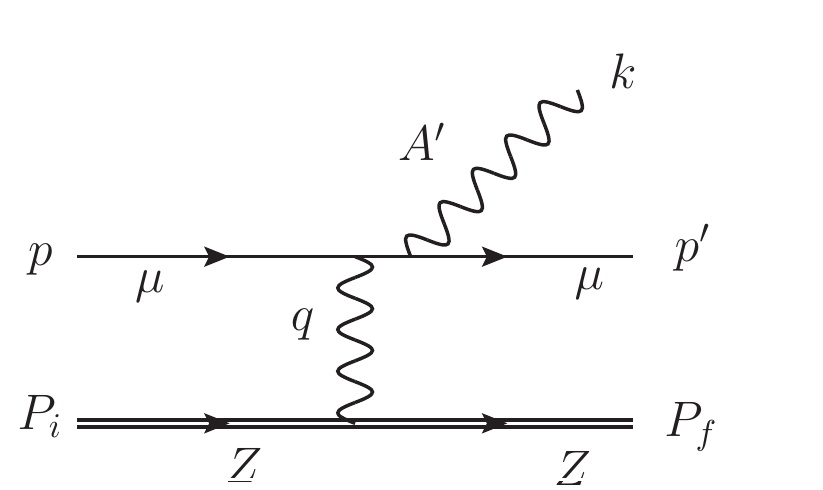}
    \end{center}
    \caption{\label{fig:mu_brems} Initial and final state bremsstrahlung production of a muon-philic vector boson $A'$ in muon-nucleus scattering.}
\end{figure}

Complementary to these two muon fixed target experiments, is the search for invisible decays of a muon-philic mediator $A'$ in rare kaon decay experiments like NA62~\cite{Krnjaic:2019rsv}. There, the $A'$ could be produced from the final state muon in the kaon decay $K\to\mu\nu$. The $A'$ can then  decay invisibly, leading again to a missing energy signature. While in principle this experiment is also capable of probing the \gmu region, NA62 will need its total planned number of $10^{13}$ collected kaons. More importantly, however, NA62 is not a background free experiment, and in order to be sensitive to \Umt as a solution to \gmu, the experimental systematics have to be further reduced~\cite{Krnjaic:2019rsv}.  We leave a detailed analysis of NA62 for future work and focus on \NAmu in this paper.

\subsection{Signals of \gmu at \NAmu}

In the \NAmu experiment, a beam of muons  with $E_0\sim 160$ GeV is dumped onto a lead target with a thickness of $L_T=40\, X_0\sim 20$ cm. The energy and momentum of the scattered particles are measured in the fiducial volume of the detector with $L_D=5$ m, consisting of the active target ECAL, a magnetic spectrometer, tracker and HCAL~\cite{Gninenko:2653581}. The production of a light muon-philic vector boson in the \NAmu target proceeds via the Bremsstrahlung process $\mu + Z \to \mu  + Z + A' $ illustrated in~\cref{fig:mu_brems}. Since the virtuality of the photon exchanged between the muon and the nucleus is quite small~\cite{Bjorken:2009mm}, the $2\to 3$ production cross section is related to real photon scattering via the Weizs\"acker-Williams approximation~\cite{Kim:1973he,Tsai:1986tx},
\begin{align}
    \dod{\sigma (\mu + Z \to \mu  + Z + A' )}{E_{A'} \,\mathrm{d}\cos\theta} &= \frac{\alpha\, \chi}{\pi} \ \frac{E_0\, \, \beta_{A'}}{1-x} \ \nonumber \\ &\times \dod{\sigma (\mu + \gamma \to \mu  +  A' )}{(p\cdot k)} \,,
\end{align}
where $\alpha$ is the fine structure constant, $\beta_{A'}=\sqrt{1-M_{A'}^2/E_0^2}$ is the boost factor of the hidden photon, $x=E_{A'}/E_0$ is the energy fraction carried away by the hidden photon, and $\chi$ denotes the effective photon flux sourced by the nucleus. The photon flux is given in terms of the electric form factor $G_2(t)$ of the nucleus as~\cite{Kim:1973he,Tsai:1973py}
\begin{equation}
    \chi = \int_{t_{\mathrm{min}}}^{t_{\mathrm{max}}} \dif t\,  \frac{t-t_{\mathrm{min}}}{t^2} \, G_2(t)\,,
\end{equation}
where $t=-q^2$ and one can approximate $t_{\mathrm{min}}=(M_{A'}^2/2E_0)^2$ and $t_{\mathrm{max}}=M_{A'}^2$. The electric form factor $G_2(t)=G_{2,\mathrm{el}}(t)+G_{2,\mathrm{in}}(t)$ has an elastic and an inelastic contribution, which are given e.g.~in Ref.~\cite{Bjorken:2009mm}.

Taking the expression for real photon scattering from Ref.~\cite{Gninenko:2014pea} and integrating over the hidden photon emission angle $\theta$, the $2\to 3$ differential cross section in the Weizs\"acker-Williams approximation is

\begin{equation}
    \od{\sigma}{x} \approx \frac{\alpha^2\,g_x^2\, \chi\, \beta_{A'}}{2 \pi^2\, (1-x)}\ \left[\frac{C_2}{V} +\frac{C_3}{2V^2} + \frac{C_4}{3V^3} \right]\,,
\end{equation}
where
\begin{align}
    C_2 &= (1-x) + (1-x)^3\,, \\
    C_3 &= -2x(1-x)^2 M_{A'}^2 - 4 m_\mu^2x(1-x)^2\,, \\
    C_4 &= 2m_\mu^4x(1-x)^3\notag\\ 
    &\quad + (1-x)^2 [4m_\mu^4  + 2m_\mu^2M_{A'}^2(x^2+(1-x)^2)]\,,
\end{align}
and
\begin{equation}
    V = M_{A'}^2 \frac{1-x}{x} +m_\mu^2 x
\end{equation}
is the virtuality of the intermediate muon. The total number of expected hidden photon events at \NAmu can then be expressed as~\cite{Chen:2017awl}
\begin{align}\label{eq:nft_full}
    N_{A'} = \mathrm{MOT} \int_{y_{\mathrm{min}}}^{y_{\mathrm{max}}} \dif y \, &n_\mathrm{atom} \int_{x_{\mathrm{min}}}^1 \dif x\, \dod{\sigma_{2\to 3}}{x} \nonumber \\ 
    &\quad\times \int_{z_{\mathrm{min}}}^{z_{\mathrm{max}}}
    \dif z\, P(z)\,,
\end{align}
where $y$ is the penetration depth of the muon in the target, $n_\mathrm{atom}$ denotes the number density of nuclei in the target material, and $x_{\mathrm{min}}$ is a cut imposed on the minimum energy fraction carried away by the hidden photon in order to suppress the background. $P(z)$ denotes the decay probability density of the hidden photon and can be written as
\begin{equation}
    P(z) = \frac{1}{\ell_{A'}} e^{-\frac{z}{\ell_{A'}}}\,,
\end{equation}
where 
\begin{equation}
\ell_{A'}= c \, \beta_{A'} \, \gamma \tau_{A'} = \frac{c\, \beta_{A'}\, E_{A'}}{M_{A'}\, \Gamma_{A'}}\,,    
\end{equation}
is the decay length. However, as long as the hidden photon is lighter than the dimuon threshold, $M_{A'}<2 m_\mu$, it will only decay invisibly (except for a very suppressed decay into $e^+e^-$ pairs) and BR(${A'}\to \mathrm{inv}) \approx 1$. In this regime, we can effectively set $z_{\mathrm{min}}=0$ and $z_{\mathrm{max}}=\infty$ so that the integration over $P(z)$ yields one.

Following Ref.~\cite{Chen:2017awl}, we note that, at the relevant energies of a few GeV to $\sim100$ GeV, the muon is a minimum ionizing particle with a rather flat stopping power $\langle \mathrm{d} E/\mathrm{d} y\rangle$ so that we can assume it to be constant~\cite{Zyla:2020zbs}.  This can be used to rewrite~\cref{eq:nft_full}, via
\begin{equation}
    y - y_{\mathrm{min}} = \frac{E_0-E_\mu}{\langle \mathrm{d} E/\mathrm{d} y\rangle}\,,
\end{equation}
in terms of the muon energy $E_\mu$ as 
\begin{equation}
    N_{A'} = \mathrm{MOT} \int_{E_{\mu, \mathrm{min}}}^{E_0} \dif E_\mu\, \frac{n_\mathrm{atom}}{\langle \mathrm{d} E/\mathrm{d} y\rangle} \int_{x_{\mathrm{min}}}^1 \dif x\, \od{\sigma_{2\to 3}}{z} \,.
\end{equation}\\
Integrating $\langle \mathrm{d}E/\mathrm{d}y\rangle$  over the length of the target $L_T\approx 40\, X_0 =20$ cm, one finds that the muon hardly loses any energy, and therefore the expression can be further simplified  to~\cite{Chen:2017awl,Kirpichnikov:2020tcf}
\begin{equation}\label{eq:ninv_simple}
    N_{A'} = \mathrm{MOT}\,  \frac{\rho N_A}{A}  \, L_T \int_{x_{\mathrm{min}}}^1 \dif x\, \od{\sigma_{2\to 3}}{x} \,,
\end{equation}
where we have expressed the number density $n_\mathrm{atom} =\rho N_A/A$ in terms of Avogadro's number $N_A$, the density $\rho$, and the mass number $A$ of the target material.

In the case of $M_{A'} <2 m_\mu$ for both \Um and \Umt, the total number of invisible $A'$ decays is given by~\cref{eq:ninv_simple}. However, in the regime where $M_{A'} \geq 2 m_\mu$, we have to be more careful since the muon-philic $A'$ can also decay visibly into muons. The invisible branching fraction is now altered, but also dimuon decays that occur outside the fiducial detector volume will be invisible. In this case, the expression for the total number of invisible events is given by

\begin{widetext}
\begin{equation}
    N_{A',\mathrm{inv}} = \mathrm{MOT}\,  \frac{\rho N_A}{A}  \, L_T \int_{x_{\mathrm{min}}}^1 \dif x\, \dod{\sigma_{2\to 3}}{x} \left(\mathrm{BR}_\mathrm{inv} + \mathrm{BR}_\mathrm{vis} \ e^{-\frac{L_\mathrm{det}}{\ell_{A'}(x)}} \right)\,,
\end{equation}
\end{widetext}
\noindent where $L_\mathrm{det}=L_T + L_D$ is the length of the target plus detector.
In the relevant coupling regime of $g_x\sim \mathcal{O}(10^{-4})$, the decay of the $A'$ is prompt at \NAmu for $M_{A'}\geq 2m_\mu$ so that almost all of them already decay within the target.

\begin{figure}[ht]
\begin{center}
\includegraphics[width=0.5\textwidth]{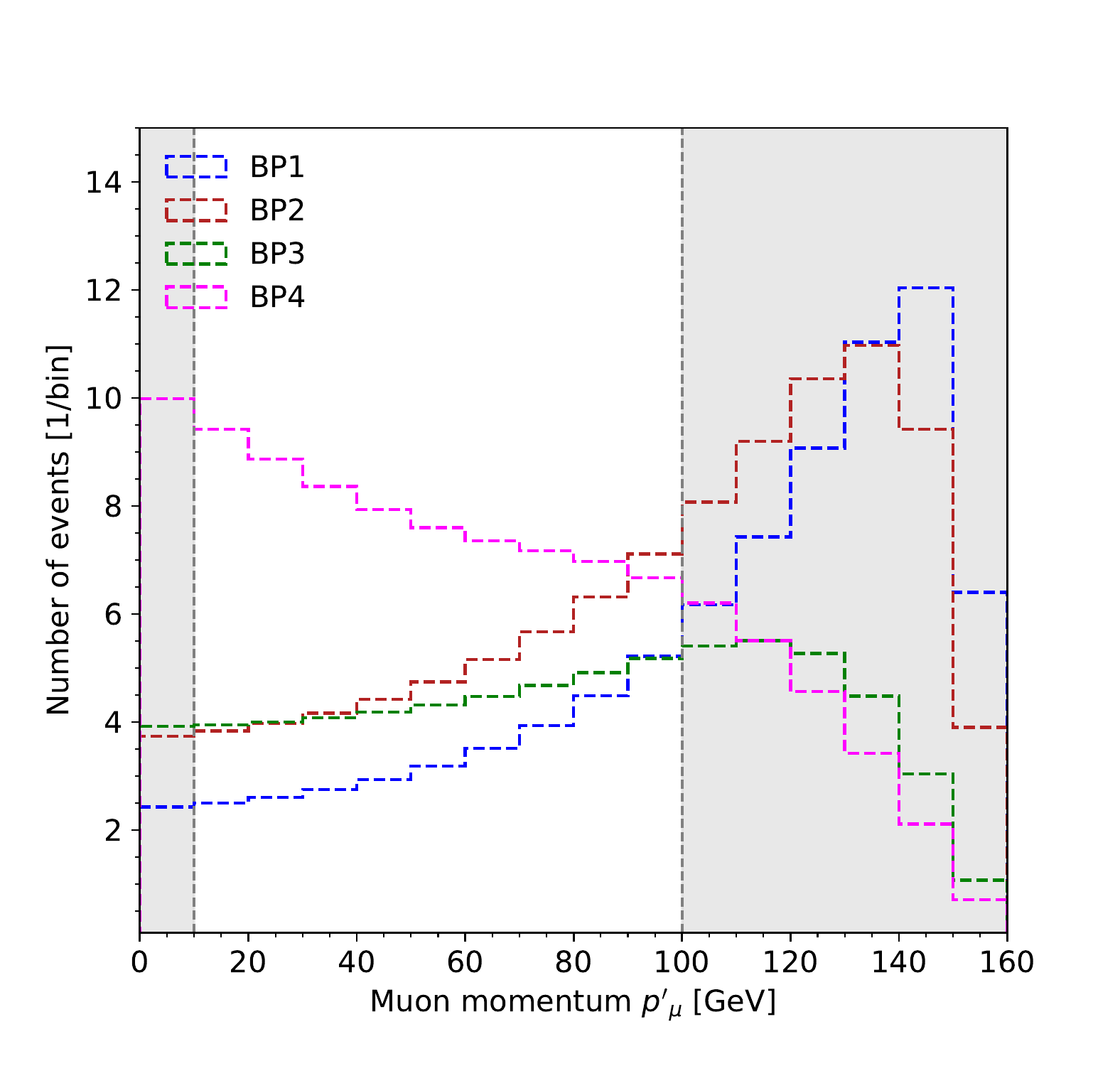}
\end{center}
\caption{Muon spectrum at \NAmu due to an invisibly decaying \Umt $A'$ signal at the benchmark points of~\cref{tab:bps}. The grey shaded areas show the experimental energy cut applied in the measurement in order to remove any backgrounds.}
\label{fig:na64_spec}
\end{figure}

The missing energy signature at \NAmu is defined by $E_{\mathrm{tot}} = E_\mathrm{ECAL} + E_\mathrm{HCAL} \lesssim 12$ GeV and a single muon carrying roughly $E'_\mu \lesssim E_0/2$ of the initial energy $E_0$. Detailed studies of the expected backgrounds have shown that imposing a maximum muon energy of $E'_{\mu}\lesssim 100$ GeV makes this search essentially background free~\cite{Gninenko:2014pea,Gninenko:2653581}. In the \NAmu proposal, the momentum resolution is given as $\sigma_p\sim1.3$ GeV and $\sigma_p\sim2.3$ GeV for a measurement of the incoming muon momentum at the beam muon station and magnetic spectrometer, respectively~\cite{Gninenko:2653581}. We assume that the resolution of the measurement of the outgoing muon momentum will be of similar order. Hence, we assume a minimum bin width of 5 GeV for the  expected signal counts as a function of the reconstructed muon energy. In order to guarantee that we have roughly $\gtrsim \mathcal{O}(3)$ events per bin for our benchmark points, we choose a more conservative binning of 10 GeV steps for our analysis. Following Ref.~\cite{Gninenko:2014pea}, we also assume a signal window of 10 GeV to 100 GeV in the muon momentum.

In~\cref{fig:na64_spec}, we show the projected muon spectra  produced by a muon-philic vector boson in the Phase-I run at \NAmu  for the benchmark points defined in~\cref{tab:bps}. These results are obtained with an assumed average signal reconstruction efficiency of $\epsilon=0.3$. Since, to our knowledge, no energy-dependent efficiency curves have been published, this is an approximation of realistic efficiencies, which lie in the range of $\epsilon=0.1 - 0.5$ for masses of 1 MeV - 1 GeV~\cite{Gninenko:2653581}.  It can be seen that all four benchmark points produce a clearly visible signal with $\mathcal{O}(1-10)$ events in each bin within the signal region.

\subsection{Parameter estimation after future detection}

Finally, with these projected signal spectra for our benchmark points at hand, we are interested in the question of how well the mass and coupling of such a muon-philic hidden photon can be reconstructed after a potential discovery at \NAmu. In order to answer this question, we perform a maximum likelihood parameter estimation.
We build our binned likelihood function based on Poisson distributed events,
\begin{equation}\label{eq:likelihood}
    \mathcal{L}(\bm{\theta}) = \prod_{i=1}^N \frac{\mu_i^{n_i}\, e^{-\mu_i}}{n_i!} \,,
\end{equation}
where $N$ is the total number of bins, $n_i$ are the observed number of events in bin $i$, and $\mu_i= s_i + b_i$ is the combined number of expected signal $s_i$ and background events $b_i$ for the model parameters $\bm{\theta}=(g_x, M_{A'})$. Since the \NAmu signal region is chosen such that the measurement is background free, we can take $b_i=0$.

Finding the set of parameters $\bm{\theta}_0$ that maximise the likelihood in~\cref{eq:likelihood}  allows us to construct the log likelihood ratio for parameter estimation according to~\cite{Feldman:1997qc},
\begin{equation}
    \ln \lambda(\bm{\theta}) = \ln \frac{\mathcal{L}(\bm{\theta})}{\mathcal{L}(\bm{\theta}_0)}\,.
\end{equation}
Since under random fluctuations in the data $n_i$ the function $-2 \ln \lambda(\bm{\theta})$ asymptotically follows a $\chi^2$ distribution with the number of degrees of freedom $n$ equal to the number of estimated parameters $\bm{\theta}$, we can reject parameters $\bm{\theta}$ at the level of $1-p$ via~\cite{Kahlhoefer:2017ddj}, 
\begin{equation}
    1 - F_{\chi^2}(n; -2\ln\lambda(\bm{\theta})) < p\,,    
\end{equation}
where $F_{\chi^2}$ denotes the cumulative $\chi^2$ distribution with $n$ degrees of freedom. Since we are estimating two parameters, the mass $M_{A'}$ and the coupling $g_x$ of the new muon-philic boson, we can exclude parameters at the 68\% CL and 95\% CL via $-2 \ln \lambda(\bm{\theta}) > 2.30$ and $-2 \ln \lambda(\bm{\theta}) > 6.18$, respectively.

\begin{figure*}[ht]
\begin{center}
\includegraphics[width=0.9\textwidth]{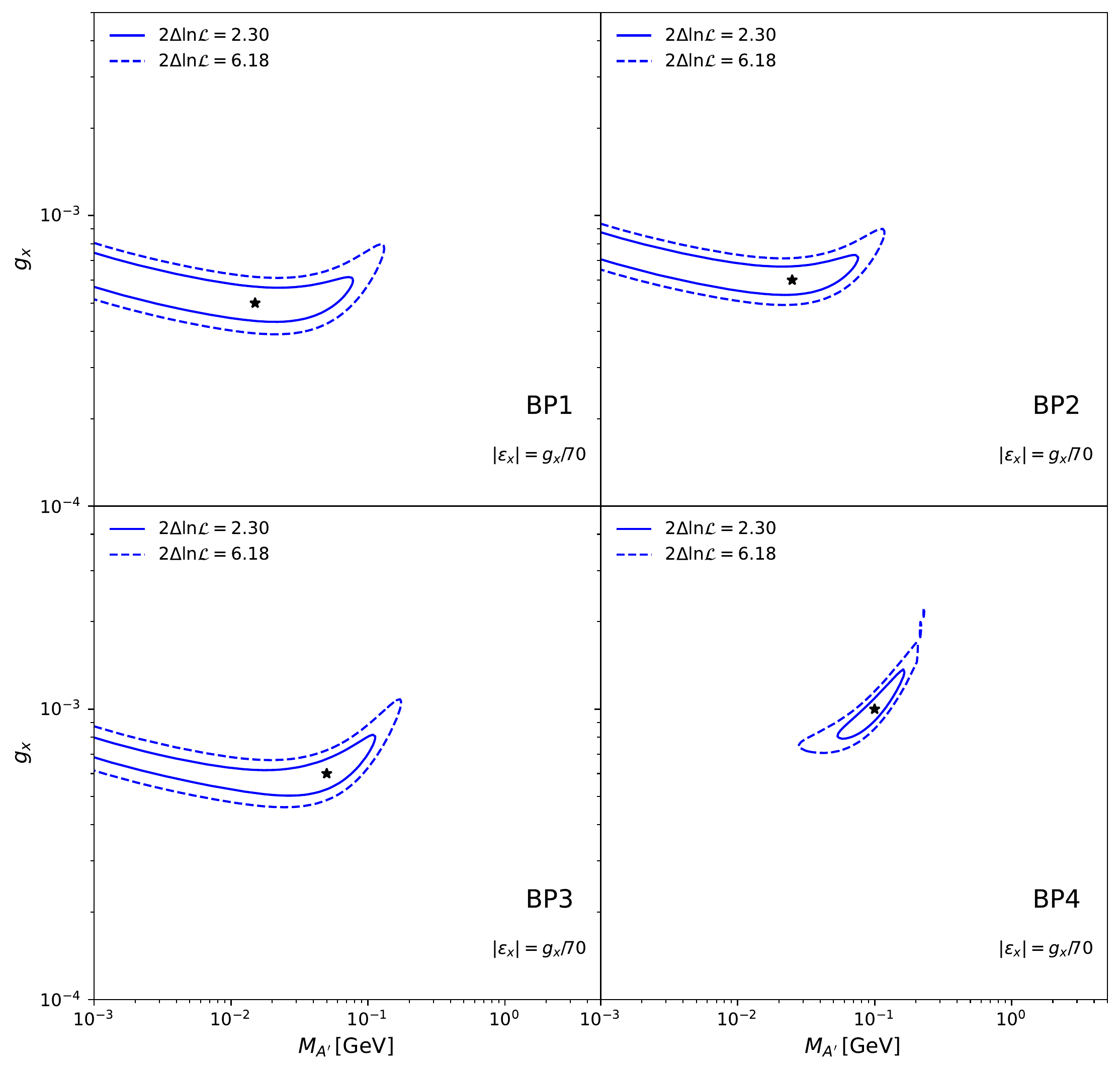}
\end{center}
\caption{Parameter reconstruction for each of the BPs of \cref{tab:bps} at NA64$\mu$. The solid (dashed) contours correspond to the 68\% CL (95\% CL). The blue contours are the parameter reconstructions for a \Um and \Umt model, which coincide for the two models for all of the BPs. The black stars mark the benchmark points.
}
\label{fig:contours_na64}
\end{figure*}

Using this method, we perform a 2D global parameter scan using the expected signal spectra of a \Umt hidden photon at the benchmark points of~\cref{tab:bps} as mock data. The resulting contours in the $(g_x, M_{A\prime})$ parameter space from both reconstructing these signals within \Umt and \Um  are shown in~\cref{fig:contours_na64}.

Let us note that \NAmu has very good sensitivity to mid-range hidden photon masses, $M_{A'}\approx 100$ MeV, in the \gmu region. This corresponds to the situation at BP4, where the $68\%$ and $95\%$ CL contours both close  around  a rather small region enclosing the true parameter point. For the lower mass BP1 - BP3, \NAmu can reconstruct the coupling $g_x$ quite well while it can only give an upper limit on the mass. This behaviour can be understood qualitatively from the spectral shapes of the signal. The spectral shape of the scattered muon is entirely fixed by the mass of the new mediator, $M_{A'}$. The coupling $g_x$ can thus be viewed as a scaling parameter that determines the overall normalisation of the spectrum. As can be seen in~\cref{fig:na64_spec}, BP4 leads to a falling spectrum (with increasing momentum) in the signal window. This flattens out slightly between $\sim60 - 90$ GeV before it falls off again more steeply towards higher momenta. This characteristic flattening is a visible feature of the spectrum within the signal window, leading to a good reconstruction of the parameters. However, BP1 - BP3 lead to a rising spectrum with their maxima lying outside the signal window at high momenta. Within the signal window, their spectral shapes are relatively similar with the characteristic peaks lying outside the signal window. This behaviour persists for smaller mass bosons, making their spectra very hard to distinguish from one another.

In general, it can be seen that the signal can be  equally well reconstructed within \Umt and \Um, and the resulting contours exactly coincide for $M_{A'}< 2\, m_\mu$ (i.e.~for all our BPs). For masses $M_{A'}\geq2\,m_\mu$, the contours for \Um and \Umt would deviate due to their invisible branching fraction of $BR_\mathrm{inv} \approx 0.33$ and  $BR_\mathrm{inv} \approx 0.5$ for \Um and \Umt above the dimuon threshold, respectively. The reason why the contours exactly coincide for $M_{A'}<2\,m_\mu$ is due to the fact that in this regime the spectra produced by a \Um or \Umt $A'$ are exactly identical since the production cross section only depends on the muon coupling and both $A'$ have an invisible branching fraction of $BR_\mathrm{inv} \approx 1$. Note that, while \NAmu in general reconstructs the coupling $g_x$ quite well, it is not at all sensitive to the kinetic mixing parameter $\epsilon_x$ (as long as $\epsilon_x \ll g_x$).  This makes it in general impossible to distinguish a potential \Um signal from \Umt by using \NAmu data alone. Hence, complementary probes of such a potential signal are needed to discriminate the two models.

\section{\cevns at spallation sources}
\label{sec:cevns}

\begin{table*}[ht!]
    \centering
    \begin{tabular}{c|c|c|c|c|c|c}
        \hline
        \hline
        Experiment & Mass [ton] & $E_{th}$ [keV$_{\rm nr}$]& NPOT  [$10^{23}/{\rm yr}$] & r & $L$ [$m$] & $\sigma_{\rm sys}$ \\
        \hline
        CENNS610 & 0.61 & $\sim 20$ & 1.5 & 0.08 & 28.4 & 8.5\% \\
        ESS10 & 0.01 & 0.1 & 2.8 & 0.3 & 20 & 5\% \\ 
        CCM & 7 & 10 & 0.177 & 0.0425 & 20 & 5\% \\
        ESS & 1 & 20 & 2.8 & 0.3 & 20 & 5\%\\
        \hline
        \hline
    \end{tabular}
    \caption{Experimental configurations for future \cevns detectors considered in our work. }
    \label{tab:experiments_cevns}
\end{table*}

New physics in the neutrino sector induces contributions to the coherent elastic scattering of neutrinos off nuclei. After the first measurement of this rare SM phenomenon by the COHERENT collaboration at a spallation source, the potential of this type of experiment to probe non-standard neutrino interactions has been thoroughly studied~\cite{Denton:2018xmq,Dutta:2019eml,Shoemaker:2021hvm,Coloma:2017ncl,Coloma:2019mbs,Flores:2020lji,Miranda:2020tif,Miranda:2020syh,Amaral:2020tga,Farzan:2018gtr,Abdullah:2018ykz,Brdar:2018qqj,AristizabalSierra:2019ykk,AristizabalSierra:2019ufd,Billard:2018jnl}. These experiments benefit from the high neutrino flux that can be achieved at spallation sources and the scalability of their detectors. In particular, a number of ton and multi-ton liquid argon detectors have been proposed that will considerably improve the sensitivity to light mediators.

At spallation sources, protons collide on a nuclear target to produce pions. These decay at rest to produce a monochromatic $29.8$~MeV muon neutrino flux ($\pi^+\to\mu^+\nu_\mu$), while the subsequent decay of the anti-muon produces a delayed beam of electron neutrinos and muon anti-neutrinos ($\mu^+\to e^+\nu_e\bar\nu_\mu$). The corresponding fluxes (see e.g., Refs.~\cite{Akimov:2018vzs,Akimov:2017ade}) read
\begin{align}
    \dod{N_{\overline{\nu} _\mu }}{E_\nu} &= \eta \frac{64E_\nu^{2}}{m_{\mu }^{3}}\left ( \frac{3}{4}-\frac{E_\nu}{m_\mu } \right ) \, ,  \nonumber
    \\
    \dod{N_{\nu _e }}{E_\nu} &= \eta \frac{192E_\nu^{2}}{m_{\mu }^{3}}\left ( \frac{1}{2}-\frac{E_\nu}{m_{\mu }} \right ) \, , 
  \nonumber\\
    \dod{N_{\nu _\mu }}{E_\nu} &=\eta\delta\left ( E_\nu-\frac{m_{\pi }^{2}-m_{\mu }^{2}}{2m_{\pi }} \right ) \, .
    \label{eq:flux_prompt}
\end{align}
The normalisation factor $\eta = rN_{\text{POT}}/4\pi L^{2}$ takes into account the number of neutrinos of each type produced from each proton on target (POT), $r$, and the baseline, $L$.

The number of expected \cevns events can then be expressed as
\begin{align}
    N_{{\rm CE}\nu{\rm NS}} = \sum_{\nu_\alpha}  N_{\mathrm{targ}} &\int_{E_{\mathrm{th}}}^{E_{R}^{\mathrm{max}}} \int_{E_{\nu}^{\mathrm{min}}}^{E_{\nu}^{\mathrm{max}}} \dod{N_{\nu _\alpha}}{E_\nu}
    \mathcal{ \epsilon } ( E_R) \nonumber  \\ 
    &\times \dod{\sigma_{\nu_{\alpha\,N}}}{E_R}
    \dif E_\nu \dif E_R \, ,
\label{eq:coherent_events}
\end{align}
where $\epsilon(E_R)$ is the energy-dependent efficiency, and the minimum and maximum neutrino energies are $E_{\nu}^{\mathrm{min}}\approx\sqrt{M_N E_R/2}$ and $E_{\nu}^{\mathrm{max}}=m_\mu/2$. The number of target nuclei is $N_{\mathrm{targ}}={M_{\mathrm{tot}}}/{M_N}$, and $M_N$ the atomic mass (we will concentrate on LAr detectors, for which we will assume 100\% isotopic abundance of $^{40}$Ar). 
The \cevns differential cross section reads~\cite{Amaral:2020tga}, 
\begin{align}
    \dod{\sigma_{\nu_{\alpha\,N}}}{E_R} &= \frac{G_F^2\, M_N}{\pi}\left(1-\frac{M_N\, E_R}{2 E_\nu^2}\right) \nonumber \\
    &\times
    \,\bigg\{ \frac{Q_{\nu N}^2}{4} \, 
    +\, \frac{g_{x}\,\epsilon_{x}\,e\, Z\ Q^x_{\nu_\alpha}\, Q_{\nu N} }{\sqrt{2}\, G_F\, (2 M_N E_R+ M_{A'}^2)} \,  \nonumber\\
    &+ \, \frac{g_{x}^2\,\epsilon_{x}^2\,e^2\, Z^2\ Q^{x^2}_{\nu_\alpha} }{2\, G_F^2\, (2 M_N E_R+ M_{A'}^2)^2}    \bigg\}\, F^2(E_R) \,,
    \label{eq:sig_numu_mur}
\end{align}
where the coherence factor for the effective neutrino-nucleus coupling via the SM $Z$-boson is defined as $Q_{\nu N} = N - (1 -4\,\sin^2\theta_W)\, Z$, and $F(E_R)$ is the Helm nuclear form factor~\cite{Helm:1956zz,Lewin:1995rx}.

Since there is no tree-level coupling between the $A^\prime$ and first generation leptons in either \Umt or \Um models, and because the $\nu_\tau$ flux is negligible (see e.g.~Ref.~\cite{Miranda:2020syh} for upper limits from oscillation parameters), the new physics contribution to $N_{{\rm CE}\nu{\rm NS}}$ only probes the coupling between $A^\prime$ and the muon sector ($\alpha=\mu$ in \cref{eq:sig_numu_mur} and \cref{eq:coherent_events}). Also, as the muon neutrino charge, $Q_{\nu_\mu}^x$, is positive and $\epsilon_x<0$, the new physics contribution from both \Umt and \Um results in a negative interference term. For the relevant part of the parameter space, this leads to a reduction in the number of expected events with respect to the SM prediction\footnote{In fact, COHERENT data on CsI, which showed a small deficit with respect to the SM expectation, has a slight preference for a $Z^\prime$ mediator coupling exclusively to the second generation \cite{Dutta:2019eml}.}, which makes these scenarios more difficult to probe.

In this section, we use data from the benchmark points defined in \cref{tab:bps} to obtain a reconstruction in the $(g_x, M_{A'})$ plane using future \cevns detectors. Since the number of events is a function of the product $g_x\,\epsilon_x$, the predictions from \Umt and \Um are, in principle, indistinguishable from each other. The reconstructed value of $g_x$ can be interpreted as different values of $g_{\mu\tau}$ and $g_\mu$ (obtained using the corresponding relation for $\epsilon_x$ from \cref{eq:epsilonmutau} and \cref{eq:epsilonmu}, respectively). However, one can expect that the combination with other types of experimental searches would help to discriminate these possibilities, as we will comment below.

We have considered four detector configurations, based on planned \cevns experiments, with parameters as detailed in \cref{tab:experiments_cevns}, inspired by the analysis of Ref.\,\cite{Miranda:2020syh}. First, we include a 610~kg (fiducial mass) extension of the current CENNS-10 LAr detector \cite {Akimov:2019xdj} at the SNS. In order to simulate the number of expected events, we have considered the same quenching factor to relate the nuclear and electron equivalent energies ($Q_F=0.246+7.8\times 10^{-4}\ E_R$, such that $E[\si{\kilo\electronvolt}\ee{}]=Q_F\, E_R$ \cite{Akimov:2020pdx}). The efficiency is expected to be similar to that of its predecessor, which dropped to 50\% at an energy of approximately 4~\si{\keV}\ee \, (equivalent to approximately $20$~\si{\keV}\nr). We have approximated it by $\epsilon = 0.5\,(1+\tanh(E_R-4~\si{\keV}\ee))$ and obtained good agreement with the spectrum predicted in Ref.~\cite{Miranda:2020syh}. Regarding the systematic uncertainty, we have fixed it at 8.5\%, as in CENNS-10. 
    
\begin{figure*}[!t]
    \begin{center}
    \includegraphics[width=0.45\textwidth]{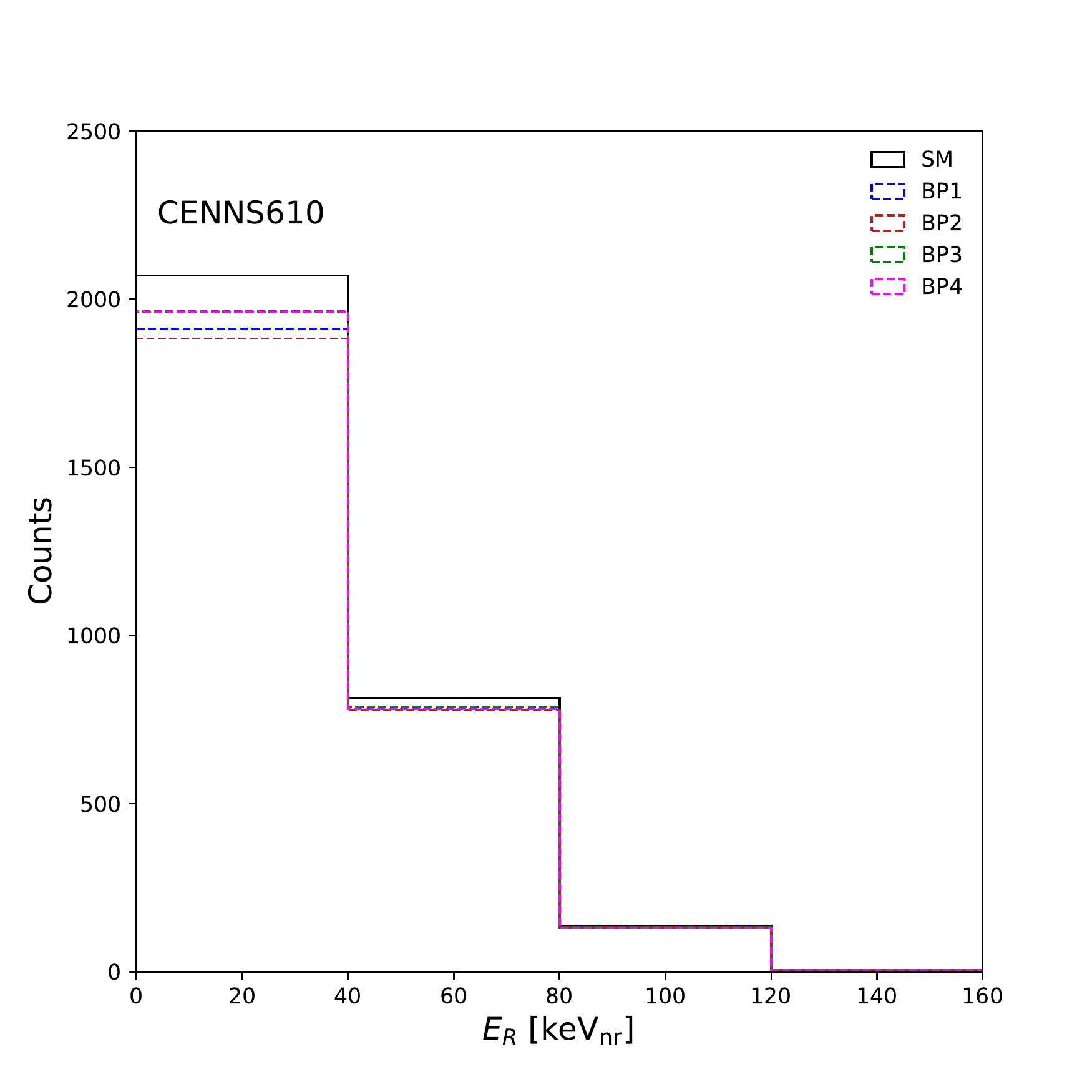}
    \includegraphics[width=0.45\textwidth]{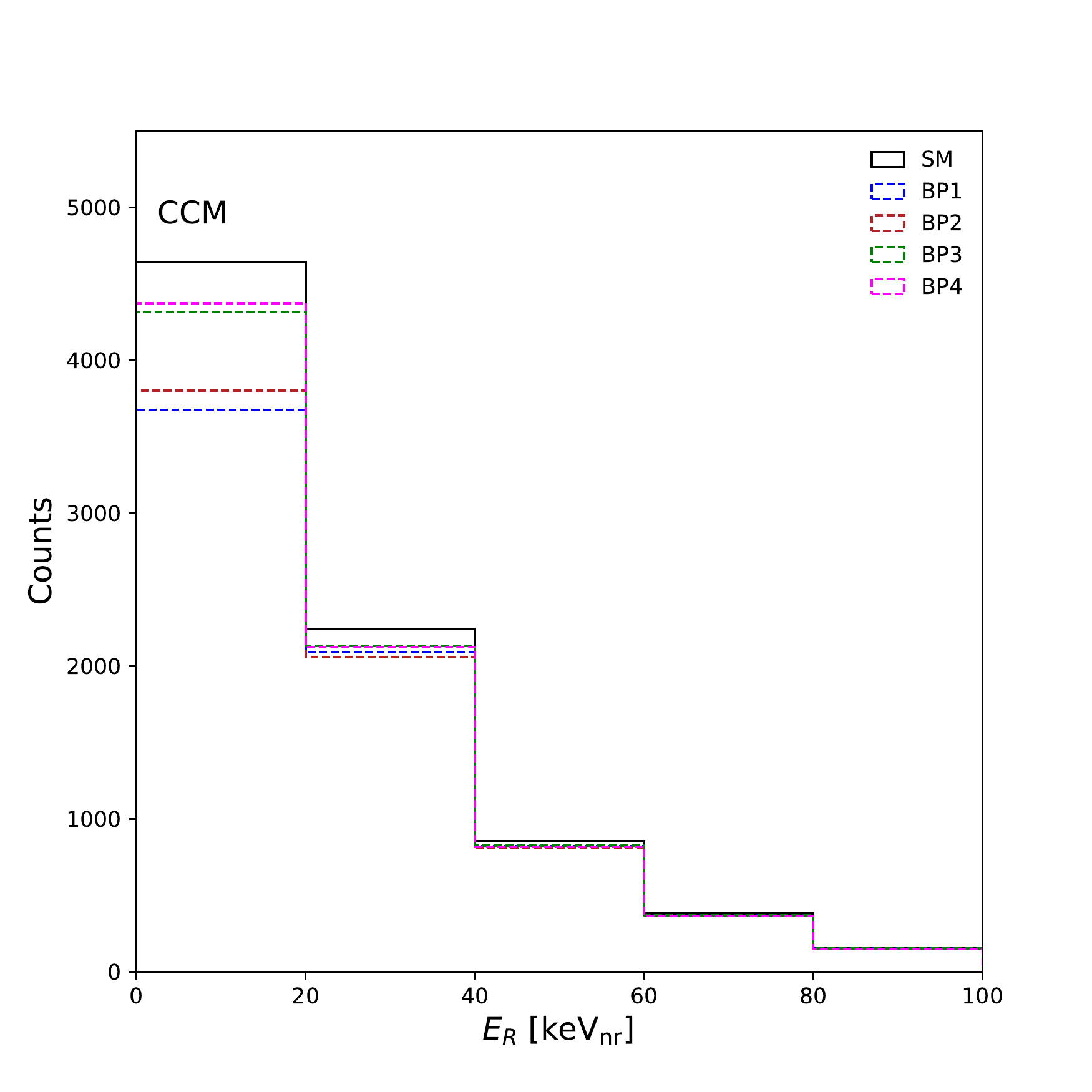}
    \includegraphics[width=0.45\textwidth]{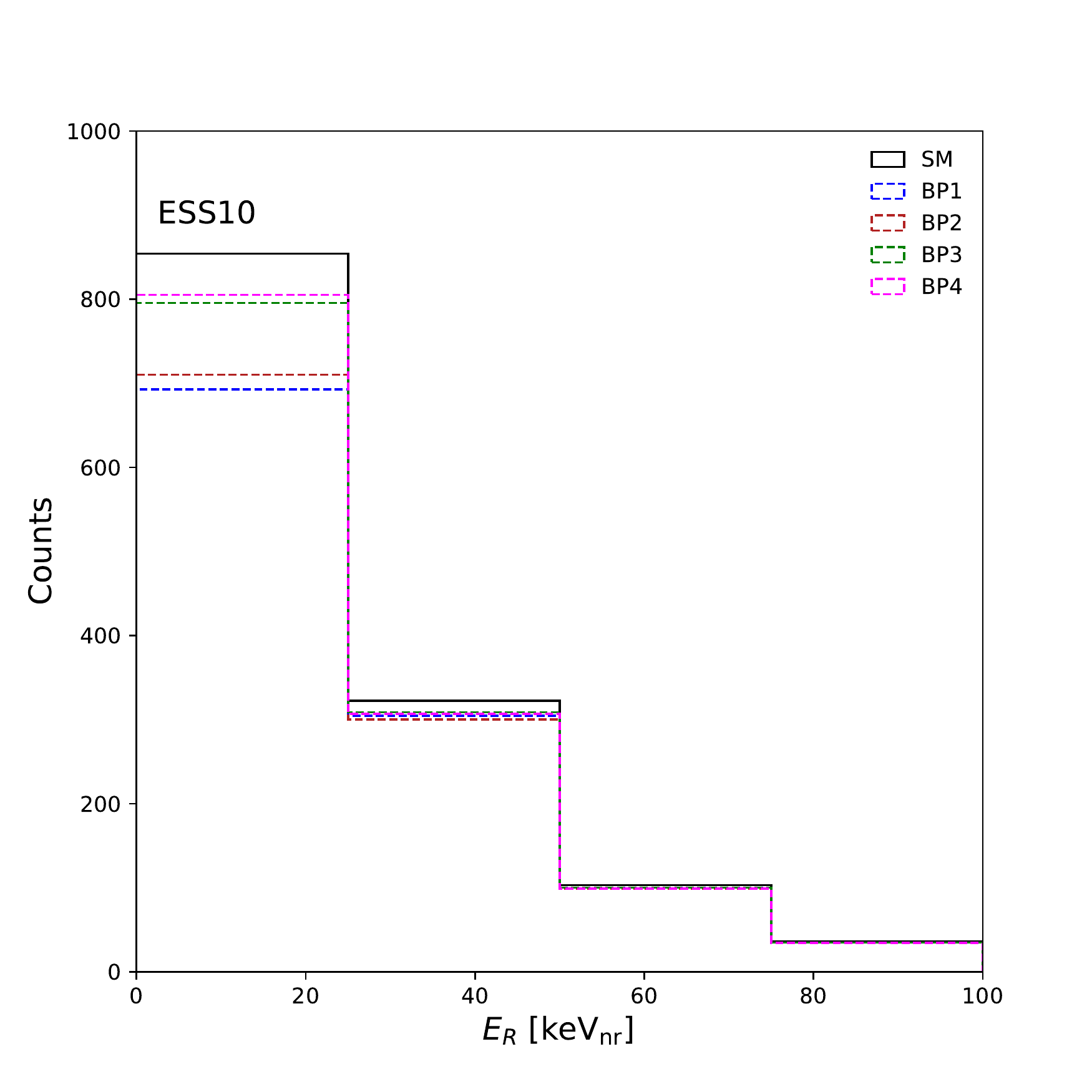}
    \includegraphics[width=0.45\textwidth]{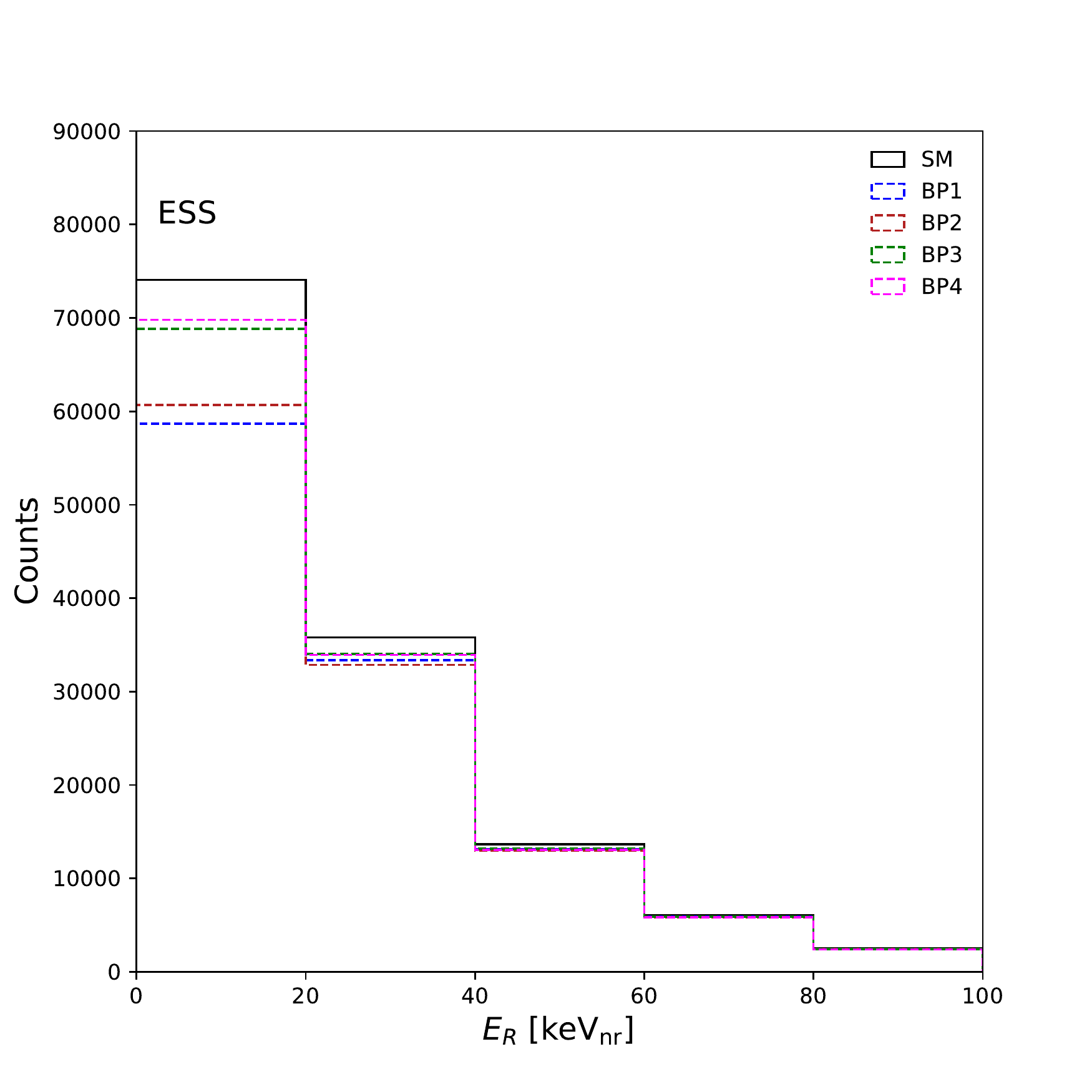}
    \end{center}
    \caption{Expected \cevns rate for each of the experimental configurations of \cref{tab:experiments_cevns}, assuming 1 year of operation. The SM prediction is shown as a solid black line, and dashed lines represent the different benchmark points of \cref{tab:bps}. For CCM, ESS10 and ESS we have only considered the events above a given energy threshold, whereas for CENNS610 we have used the efficiency of its predecessor CENSS-10 LAr. }
    \label{fig:ccm_bps}
\end{figure*}

Based on the projected characteristics of CCM \cite{ccm}, we have considered a fiducial mass of 7 tons of LAr. CCM is planned to run for a total of 2.5 years at Lujan in a near position ($L=20$~m) and a far position ($L=40$~m) configuration. We have only used data from the near position run (as the expected number of \cevns events is approximately four times larger), assuming a total operation time of 1~yr in this configuration.  Regarding the ESS \cite{Baxter:2019mcx}, we proceed as in Ref.~\cite{Miranda:2020syh} and consider two different setups: a small (10~kg) but extremely sensitive ($E_{\rm th}=0.1$~keV) detector, and a large one (1~ton) with the same threshold energy as CCM and CENNS. For both configurations, the baseline is $L=20$~m and we assume 1~yr of operation.

\begin{figure*}[!t]
    \begin{center}
    \includegraphics[width=0.9\textwidth]{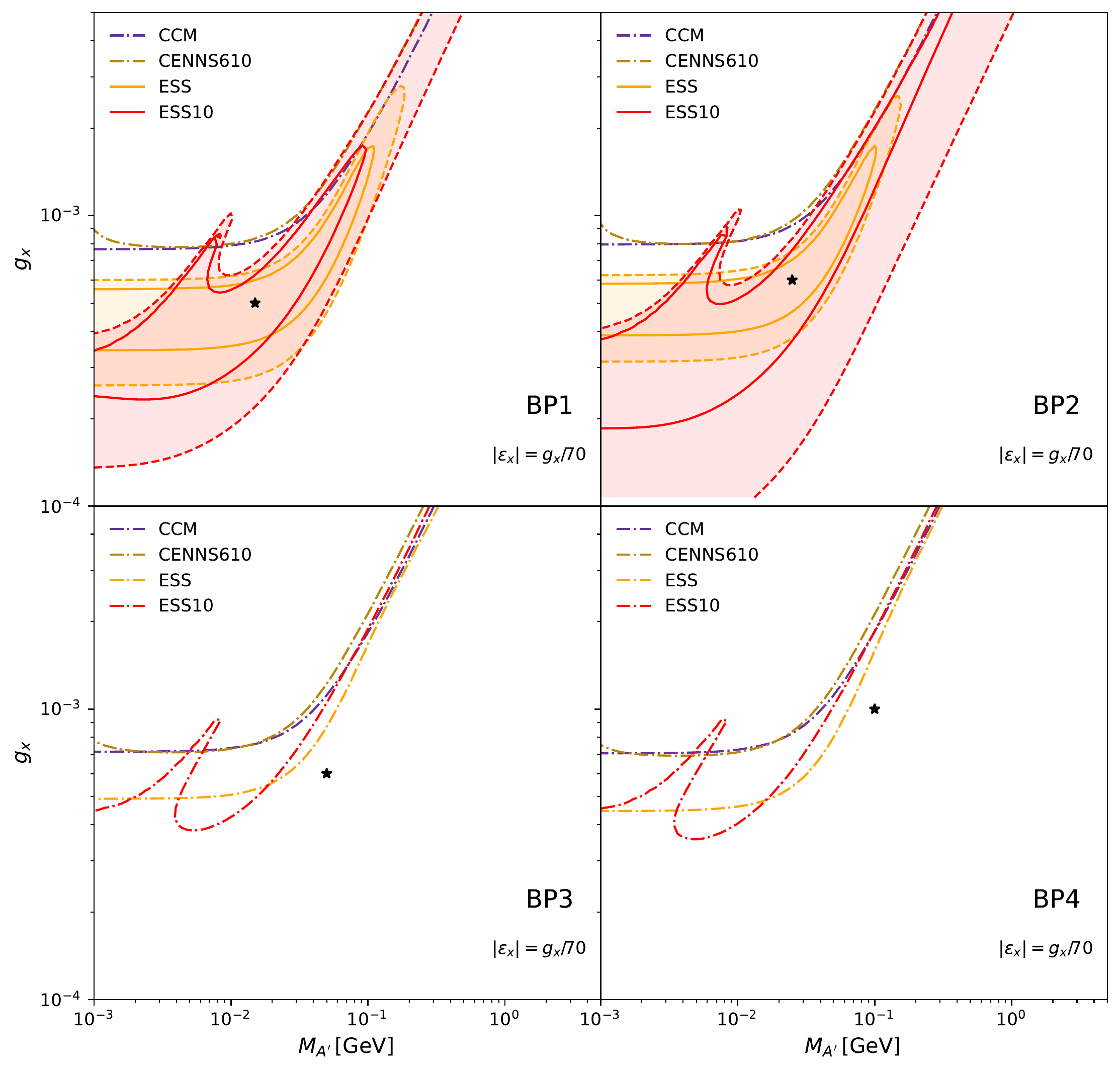}
    \end{center}
    \caption{Parameter reconstruction for each of the BPs of \cref{tab:bps} using \cevns data from the experiments of \cref{tab:experiments_cevns} and fixing $\epsilon_x=-g_x/70$. The solid (dashed) contours correspond to the 68\% CL (95\% CL) and the star represents the position of each benchmark point. The dot-dashed lines denote the 90\% CL upper limit on the coupling $g_x$ in the case no reconstruction is possible. }
    \label{fig:contours_cevns}
\end{figure*}

For both CCM and ESS, we consider 100\% efficiency and we do not include any quenching factor. Regarding the threshold, we will assume a sharp cut at $E_{\mathrm{th}} = 20~\si{\keV}\nr$ (this value is similar to the energy at which the efficiency of CENNS-10 LAr drops to 50\%).

An important background for this kind of search is due to beam related neutrons (BRN), which for CENNS-10 represented approximately 10\% of the signal events \cite{Akimov:2019xdj}. Since we ignore how CCM and ESS would perform in this respect, we have assumed $N_{\mathrm{bkg}} = 10\% N_{\mathrm{SM}}$. We will consider four energy bins across the energy window of $20-100$~keV. Having such a modest measurement of the energy spectrum allows for some limited reconstruction of the mass of the mediator in the event of an observation, but most importantly, it helps reducing the effect of the normalisation systematic error.

\begin{figure*}[!ht]
    \begin{center}
    \includegraphics[width=0.9\textwidth]{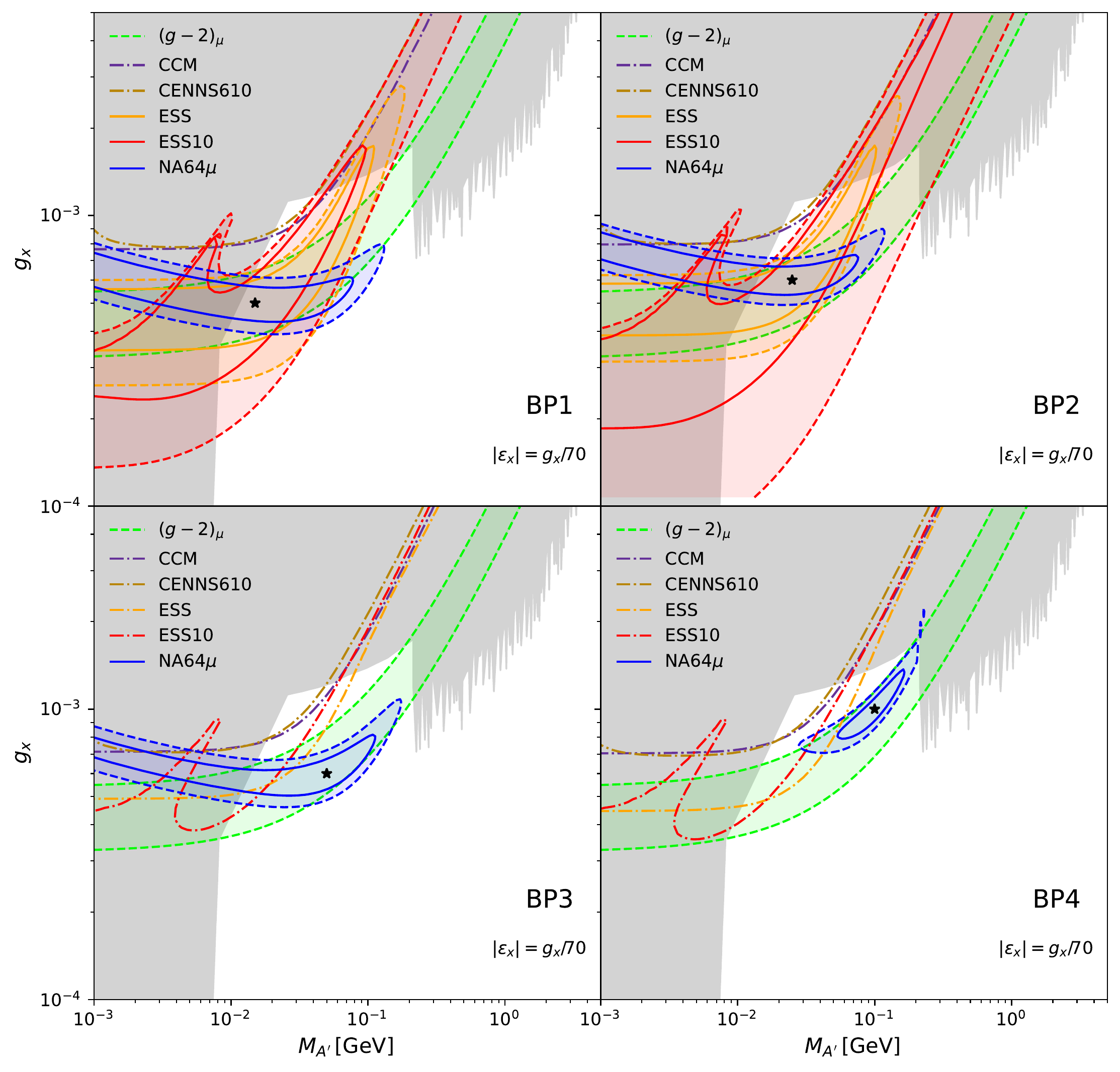}
    \end{center}
    \caption{The same as in \cref{fig:contours_cevns}, but adding data from \NAmu and interpreting the results in a \Umt model. The green band shows the region favoured by the recent \gmu determination.}
    \label{fig:cevns+na64_mutau}
\end{figure*}

We have used a $\chi^2$ test to compare the observed number of events in the $i$-th energy bin, $N_{\rm obs}^i$, with the theoretical prediction for a given point in the parameter space, $N_{\rm th}^i(g_{x},M_{A'})$, 


\begin{widetext}
\begin{equation}
    \chi^2 (g_{x},M_{A'}) = {\rm min}_a \left[
    \sum_i\frac{\left(N_{\rm obs}^i-N_{\rm th}^i(g_{x},M_{A'})[1+a]\right)^2}{(\sigma^i_{\rm stat})^2}+\left(\frac{a}{\sigma_{\rm sys}}\right)^2\right]\ .
\end{equation}
\end{widetext}

\noindent The statistical uncertainty is defined as $$\sigma^i_{\rm stat} = \sqrt{N_{\rm obs}^i+N_{\mathrm{bkg}}^i}\,,$$ where $N_{\mathrm{bkg}}^i=N_{\mathrm{SM}}^i/10$ (and is therefore the same for all benchmark points). There is a systematic uncertainty in the total normalisation, represented by a nuisance parameter, $a$, which we take as 5\%. We consider this is a realistic goal, based on the 8.5\% systematic error achieved in CENNS-10 LAr.

The energy spectra for the four benchmark points of \cref{tab:bps} are represented in \cref{fig:ccm_bps}, where the SM prediction is shown as a black solid line for comparison. As mentioned above, all BPs show a deficit of events with respect to the SM rate. The difference is much more prominent at low energies, which is an excellent motivation to reduce the experimental threshold in this kind of experiment.

\begin{figure*}[!ht]
    \begin{center}
    \includegraphics[width=0.9\textwidth]{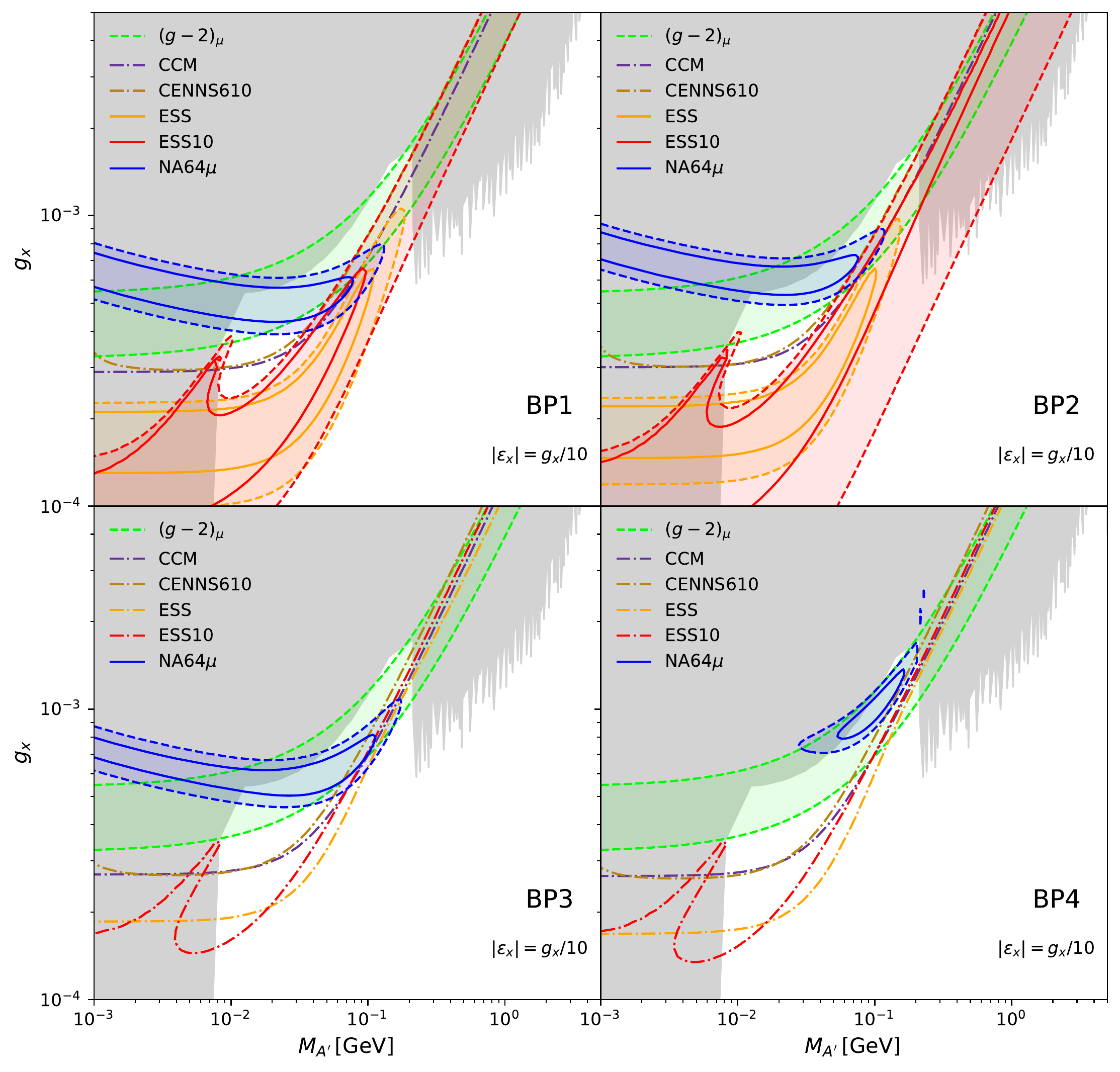}
    \end{center}
    \caption{The same as in \cref{fig:cevns+na64_mutau}, but for $\epsilon_x=-g_x/10$.}
    \label{fig:cevns+na64_mu}
\end{figure*}

\cref{fig:contours_cevns} shows the parameter reconstruction in the $(g_x,M_{A'})$ plane for each of the benchmark points of \cref{tab:bps} and for the four experimental configurations of \cref{tab:experiments_cevns}. For this plot, we have fixed $\epsilon_x = -g_x/70$, which coincides with the relation expected for a \Umt model. The true position of each benchmark point is shown by means of a star. The sensitivity of these detectors is optimal for mediator masses below approximately 50~MeV, corresponding to the maximum momentum exchange from the incoming neutrino flux $2\,M_N\,E_R$. Above this mass, the mediator mass dominates the propagator of the new physics contributions and the scattering cross section \cref{eq:sig_numu_mur} decreases as $g_x^2\,M_{A'}^{-2}$ (as long as the interference term dominates). This means that only the low-mass window of the area compatible with \gmu can be probed: of all the benchmark points in \cref{tab:bps}, BP1 and BP2 would be observed in the future ESS. For BP3 and BP4, we obtain upper bounds on the coupling as a function of the mediator mass.

In particular, the upcoming CENSS610 and CCM experiments would be unable to disentangle the SM \cevns flux from a new physics signal, resulting in upper bounds on the parameter space that reach couplings of the order of $g_x\sim 7\times 10^{-4}$ (for CCM, this is in agreement with the results from Ref.~\cite{Banerjee:2021laz}). The predictions for ESS are more optimistic. The case of ESS10 perfectly illustrates the great benefit of lowering the experimental threshold: a small but extremely sensitive configuration would be able to observe the nuclear recoil spectrum where the new-physics contribution from light mediators is maximal (see the lower left panel in \cref{fig:ccm_bps}). Not only does this allow the new physics prediction to be disentangled from that of the SM (thereby producing closed contours in $g_x$), but also some sensitivity to the mediator mass is obtained (and upper bounds are found for $M_{A'}$). The bump at low mediator masses in the ESS10 lines corresponds to the crossover of the contribution from the interference term and the pure BSM one (only observable at low energies). The 1 ton configuration of ESS would perform slightly better at larger masses, and could confirm observation for BP1 and BP2 (providing closed contours in the coupling $g_x$ but a poor reconstruction of the mediator mass).

It should be emphasised that the results of \cref{fig:contours_cevns} can be interpreted in both the \Umt or \Um model. Experiments looking for \cevns at spallation sources, on their own, would be unable to measure the kinetic mixing parameter $\epsilon_x$.

\subsection{Combination with \NAmu}
\label{sec:cevns+na64}

The combination of results from \NAmu and \cevns experiments at spallation sources can potentially shed some light on the value of the kinetic mixing parameter. As explained in \cref{sec:beam_dump}, \NAmu is only sensitive to $g_x$, whereas \cevns probes $g_x\epsilon_x$. Thus, the combination of results from both sources could be used to infer the value of $\epsilon_x$. In this section we illustrate this complementarity, showing that tension in the reconstructed regions arises if the wrong assumption is made for $\epsilon_x$.

First, \cref{fig:cevns+na64_mutau} shows the combined results from future spallation sources and \NAmu, assuming $\epsilon_x=-g_x/70$. This coincides with the prediction in a \Umt model and with the assumption that we made in generating the benchmark points. The evidence strengthens for BP1 and BP2, since these benchmark points can be probed by both types of techniques. Even if spallation source experiments do not observe new physics (as is the case of BP3 and BP4) the resulting bounds can narrow down the region compatible  with \NAmu, providing a better measurement of the mediator mass.

In contrast, \cref{fig:cevns+na64_mu} assumes $\epsilon_x=-g_x/10$ in the reconstruction, a value that could come from a particular realisation of a \Um model. As expected, the reconstructed areas show increasing tension. For example, for BP1, there is a small overlap of the 95\% CL contours, which shrinks for BP2. In contrast, for BP3 and BP4 almost the whole area compatible with \NAmu is excluded, indicating a wrong assumption for $\epsilon_x$.

Measuring $\epsilon_x$ does not necessarily imply discriminating \Umt and \Um. In particular, if $\epsilon_x$ is shown to be inconsistent with $-g_x/70$, then the minimal \Umt model would be ruled out as an explanation for $(g-2)_\mu$. However, if $\epsilon_x$ is found to be compatible with $-g_x/70$, this could correspond to either \Umt or \Um. The only way to tell these constructions apart would be to explore the couplings to the tau sector. As we will argue in the next section, this could be done by observing solar neutrinos in direct detection experiments.

\section{Direct detection experiments}
\label{sec:direct}

The expected differential rate of solar neutrino-electron or \cevns scattering events at direct dark matter detection experiments can be written as 
\begin{equation}
    \label{eq:dr_dd}
    \od{R}{E_R}= n_T  \sum_{\nu_\alpha}\int_{E_\nu^\mathrm{min}} \od{\phi_{\nu_e}}{E_\nu} \ P(\nu_e \rightarrow \nu_\alpha)\  \od{\sigma_{\nu_{\alpha\,T}}}{ E_R} \ \dif E_\nu\, ,
\end{equation}
where $E_{\nu}^{\mathrm{min}} = (E_R + \sqrt{E_R^2 + 2 m_T E_R} )/2$  is the minimum neutrino energy to produce a nuclear recoil or electronic recoil of energy $E_R$, $m_T$ is the mass of the target, and $n_T$ is the total number of targets (electrons or nuclei) per unit mass. Regarding the solar neutrino fluxes, $\mathrm{d} \phi_{\nu_e}/{\mathrm{d} E_\nu}$, we will consider a high metallicity scenario, which increases the flux of neutrinos in the $pp$-chain by up to 10\%. For the case of $^8\mathrm{B}$ neutrinos, this corresponds to a total flux of $5.46\times10^{-6}\,\si{\cm^{-2}\s^{-1}}$ \cite{Vinyoles:2016djt}. The energy-dependent oscillation probability $P(\nu_e \rightarrow \nu_\alpha)$ is taken from Fig.~7 of Ref.~\cite{Amaral:2020tga}.

The differential cross section of the elastic scattering of neutrinos off nuclei is given in \cref{eq:sig_numu_mur}, and the corresponding expression for scattering off electrons for $\alpha = \mu, \tau$ reads

\begin{widetext}
\begin{align}\label{eq:sig_el_nsi}
    \dod{\sigma_{\nu_{\alpha\,e}}}{E_R} =   &\frac{2\, G_F^2\,m_e}{\pi} \ \Bigg\{ \,  \left[ {g^e_L}^2+{g^e_R}^2\left(1-\frac{E_R}{E_\nu}\right)^2 - g^e_L\,g^e_R \frac{m_e \, E_R}{E_\nu^2} \right]  \notag \\
    &+ \frac{g_{x}\, \epsilon_{x}\, e \,  Q^x_{\nu_\alpha}}{\sqrt{2}\, G_F (2E_R\, m_e + M_{A'}^2)} \left[ (g^e_L+g^e_R)\left(1-\frac{m_e\, E_R}{2E^2_\nu}\right) - g^e_R \frac{ E_R}{E_\nu}\left(2- \frac{E_R}{E_\nu}\right) \right]  \notag \\
    &+  \frac{g_{x}^2\, \epsilon^2_{x} \, e^2 \,  Q^{x^2}_{\nu_\alpha}}{\,4\, G_F^2 (2E_R\, m_e + M_{A'}^2)^2} \left[  1 - \frac{ E_R}{E_\nu} \left(1 - \frac{E_R-m_e}{  2\, E_\nu}\right) \right]\, \Bigg\} \,,
\end{align}
\end{widetext}
\noindent where $g_L^e = \sin^2\theta_W - 1/2$ and $g_R^e = \sin^2\theta_W$ are the couplings of the $Z$-boson to the electron. The SM expression for $\alpha=e$ is obtained by setting $\epsilon_x=0$ and replacing $g_L^e \rightarrow 1 + g_L^e$ in the above expression. For more details, see Ref.~\cite{Amaral:2020tga}.
The first term in~\cref{eq:sig_el_nsi} corresponds to the pure SM, the second to the interference, and the last one to the pure BSM contribution. The sign of the interference term critically depends on the charge  $Q^x_{\nu_\alpha}$ of the scattered neutrino $\nu_\alpha$, which is positive for muon and negative for tau neutrinos. For electron recoils, the pure BSM term dominates the scattering cross-section at couplings relevant for explaining \gmu, and therefore we expect an increase in the number of events for both \Umt and \Um.

When calculating the expected number of scattering events in direct detection experiments, we include detector-specific effects, such as detector efficiencies and resolutions. Efficiencies $\epsilon(E_R)$ are folded into \cref{eq:dr_dd} before the theoretical differential rate is convoluted with a Gaussian resolution function with energy-dependent width, $\sigma(E_R)$. This results in an expected observed count of 
\begin{align}
    \label{eq:n_dd}
    N = \varepsilon\int_{E_\mathrm{th}}^{E_\mathrm{max}}&\Bigg(\int_0^\infty \dod{R}{E'}\epsilon(E')  \nonumber \\ 
    &\times \frac{1}{\sigma(E')\sqrt{2\pi}}e^{-\frac{\left(E_R - E'\right)^2}{2\sigma^2(E')}}\,\dif E'\Bigg) \dif E_R\,,
\end{align}
where $\varepsilon$ is the exposure of the experiment, and $E_\mathrm{th}$ is its energy threshold.

An important difference with respect to the physics at \NAmu and \cevns experiments at spallation centres is that there is now an incoming flux of tau neutrinos. In fact, given the neutrino oscillation probabilities of Ref.~\cite{Amaral:2020tga}, the flux of $\nu_\tau$ is larger than that of $\nu_\mu$ at the energies relevant for the $^8\mathrm{B}$ and $hep$ solar fluxes. This is extremely relevant for discriminating new physics models. In particular, the \Umt contribution for both NR and ER is positive with respect to the SM. In contrast, the \Um model with $\epsilon_\mu=-g_\mu/70$ leads to a negative contribution to NR (from the dominant interference term) and a positive one for ER (from the pure BSM term). Therefore, an observation of NR in direct detection experiments would unequivocally point to either a \Umt or \Um model.

This is by no means an easy task: although the flux of solar neutrinos is much larger than that of atmospheric ones, their energies only reach approximately 15~MeV (for $^8\mathrm{B}$). These lead to $\sim$keV nuclear recoils, right at the threshold of current liquid xenon detectors. On the other hand, although solid-state detectors benefit from a much lower threshold, their planned target size makes them unlikely to observe enough events from solar neutrinos in the near future. Thus, in our analysis, we will concentrate on planned LXe detectors and will base our experimental configurations on the upcoming multi-ton LZ \cite{Mount:2017qzi} and XENONnT \cite{Aprile:2020vtw} detectors and on the projected DARWIN observatory \cite{Aalbers:2016jon}. The proposed exposures are 15.34, 20 and 200 $\textrm{ton}\cdot\textrm{yr}$ for LZ, XENONnT and DARWIN, respectively.

For LZ, we have taken the efficiency functions given in \cite{Akerib:2018lyp}, the resolution fit given by LUX in \cite{Akerib:2016qlr}, and ER backgrounds from \cite{Akerib:2021qbs}. For XENONnT, the NR and ER efficiency functions have been taken from \cite{Aprile:2018dbl} and \cite{Aprile:2020tmw}, respectively,  the resolution function has been read from \cite{Aprile:2020tmw}, and the ER backgrounds have been taken from \cite{Aprile:2020vtw}. Finally, for DARWIN, we have assumed the same efficiency and resolution functions as for XENONnT, used the lower ER background predictions given by \cite{Baudis:2013qla}, and assumed a flat background rate for all background components except for the double-$\beta$ decay of $^{136}$Xe. For all experiments, we have taken the nominal energy thresholds to be at the energies where their (event-specific) efficiencies reach 50\%. This corresponds to LZ thresholds of $3.8\,\knr$ and $1.5\,\kee$, and XENONnT and DARWIN thresholds of $5.7\,\knr$ and $1.5\,\kee$.

We have performed three different types of analyses: an NR-only analysis (based on the discrimination of nuclear and electron recoils to reduce background), an ER-only analysis, and a combined NR + ER analysis. For the NR-only analysis, we assume a 50\% acceptance cut above the energy threshold after rejecting ER events and treat this as a background-free analysis. For the ER-only analysis, we assume 99.5\% ER/NR discrimination and include experimentally-relevant ER backgrounds. For the NR + ER analysis, we combine both types of events, interpreting all events as ER events, with the ER background included but without cuts. In each analysis, the appropriate NR/ER efficiency functions are folded into the differential rate, with the energy resolution taken at the appropriate energy scale.


\begin{figure*}[!t]
    \begin{center}
    \includegraphics[width=0.47\textwidth]{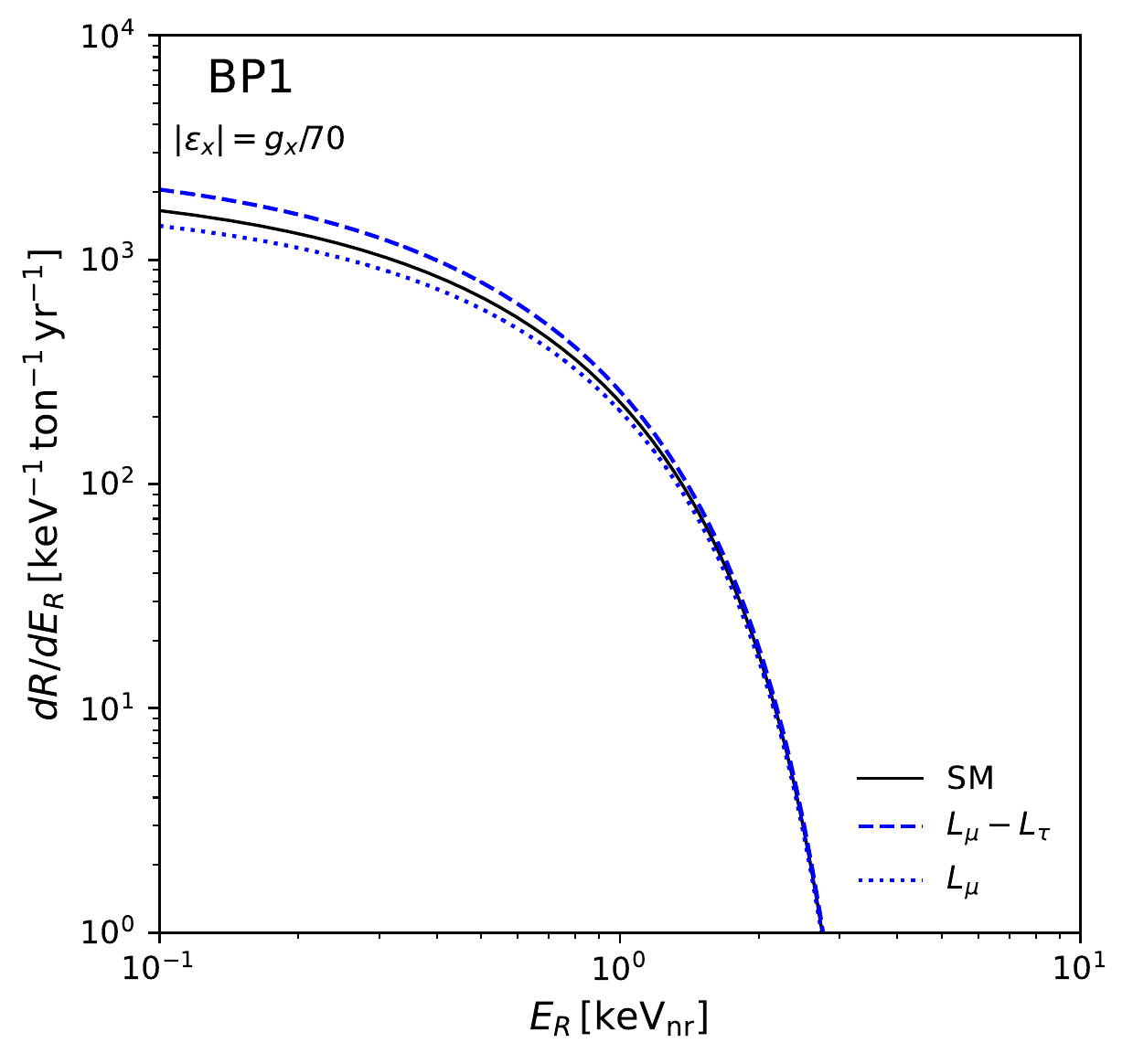}
    \includegraphics[width=0.495\textwidth]{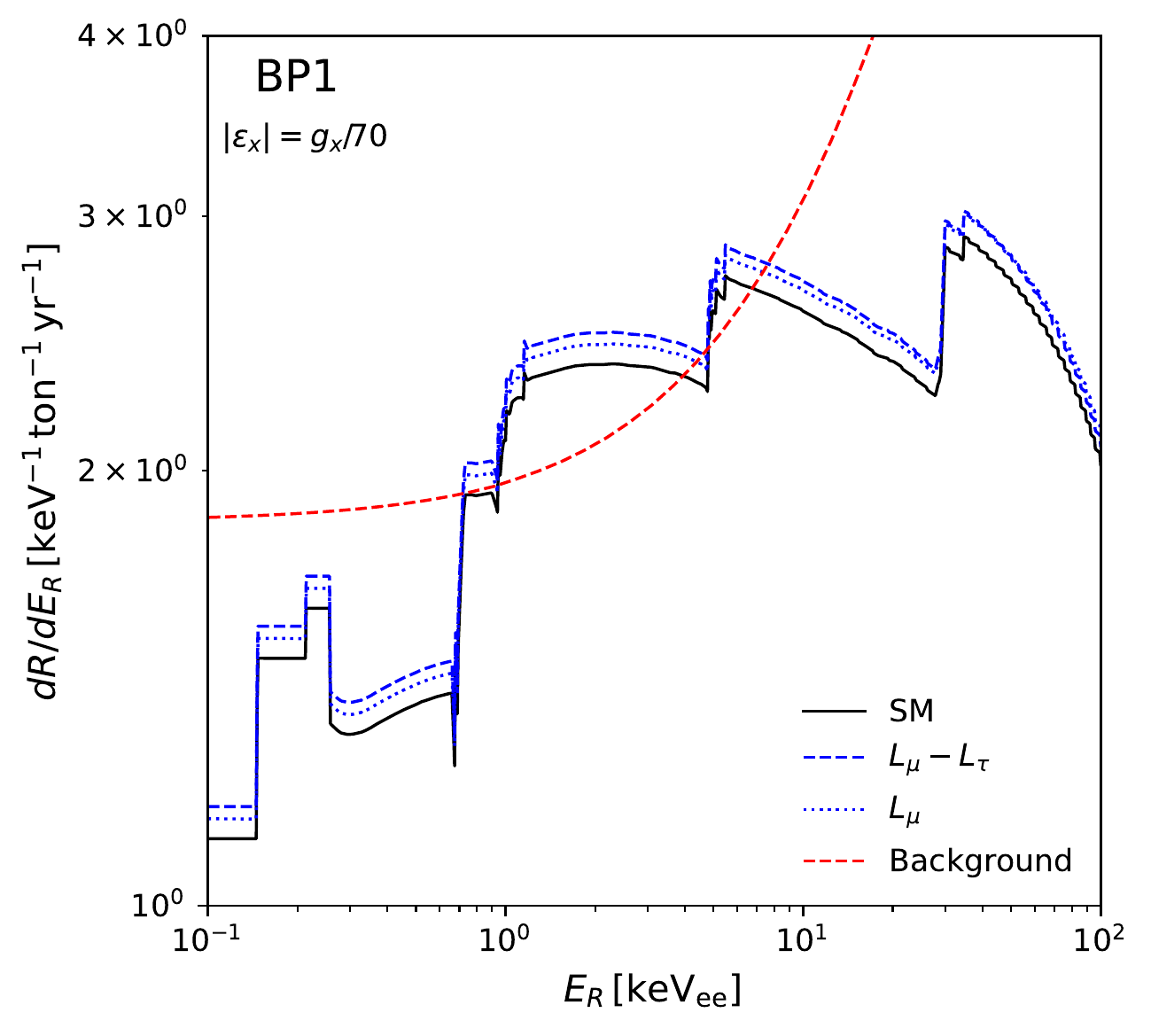}
    \end{center}
    \caption{\label{fig:xe_spec_lz} (Left) Nuclear-recoil spectrum for \cevns for a generic Xe detector. The SM prediction is shown in black, while the BSM prediction for BP1 is shown in blue for \Umt (dashed) and \Um (dotted). (Right) Electron recoil spectrum for a generic Xe detector. A rough energy-independent scaling has been applied within the $0.25 - 30$~keV energy window to account for the RRPA. We also show in this plot the relevant background by the red dashed line, based on the predictions for DARWIN given in Ref.~\cite{Baudis:2013qla}.}
\end{figure*}

The left panel of \cref{fig:xe_spec_lz} shows the predicted differential rate for nuclear recoils in a generic xenon detector. The NR signal from solar neutrinos from the $^8\mathrm{B}$ flux falls abruptly at approximately $1-2$~keV. This makes their observation very challenging, forcing experiments to achieve a very low energy threshold. For the nominal value of the threshold of the planned LZ and XENONnT detectors (of the order of 3~keV), only a handful of these neutrinos can actually be observed (mostly due to the effect of the resolution near the threshold). In order to exploit this signal one must therefore achieve lower experimental thresholds. The figure also illustrates how the \Umt model predicts an excess of events with respect to the SM, whereas \Um would lead to a reduction of the rate. Given the importance of nuclear recoils in discriminating between \Um and \Umt, we take the liberty to vary the threshold, exploring their effect with regards to discovery potential.

The right panel of \cref{fig:xe_spec_lz} corresponds to the expected electron recoil rate for BP1, compared to the SM prediction. For comparison, we also show the expected background, using the projection for DARWIN as guideline. The electron background is dominated by $^{136}$Xe double-$\beta$ decay at high energies. For the energy range considered in the analysis, one must take into account the fact that electrons are originally bound to the atom. This results in a series of steps in the spectrum, which in a first approximation, correspond to the different ionisation energies and effectively reduces the number of electrons available at low energies. A more careful treatment in terms of a relativistic random phase approximation (RRPA) leads to a further suppression in the $0.25 - 30~\kee$ window according to Fig.~2 of Ref.\cite{Chen:2016eab}, which we have implemented using an energy-dependent scaling\footnote{In principle, one should also take into account how the RRPA result changes in the presence of new light mediators, but this is beyond the scope of our work.} . Note that, below $0.25~\kee$, we have reverted back to the step-function approximation, which acts as an upper bound to the expected rate in the absence of numerical solutions to the RRPA at low energies \cite{Chen:2016eab}.

\begin{figure*}[!t]
    \begin{center}
    \includegraphics[width=0.9\textwidth]{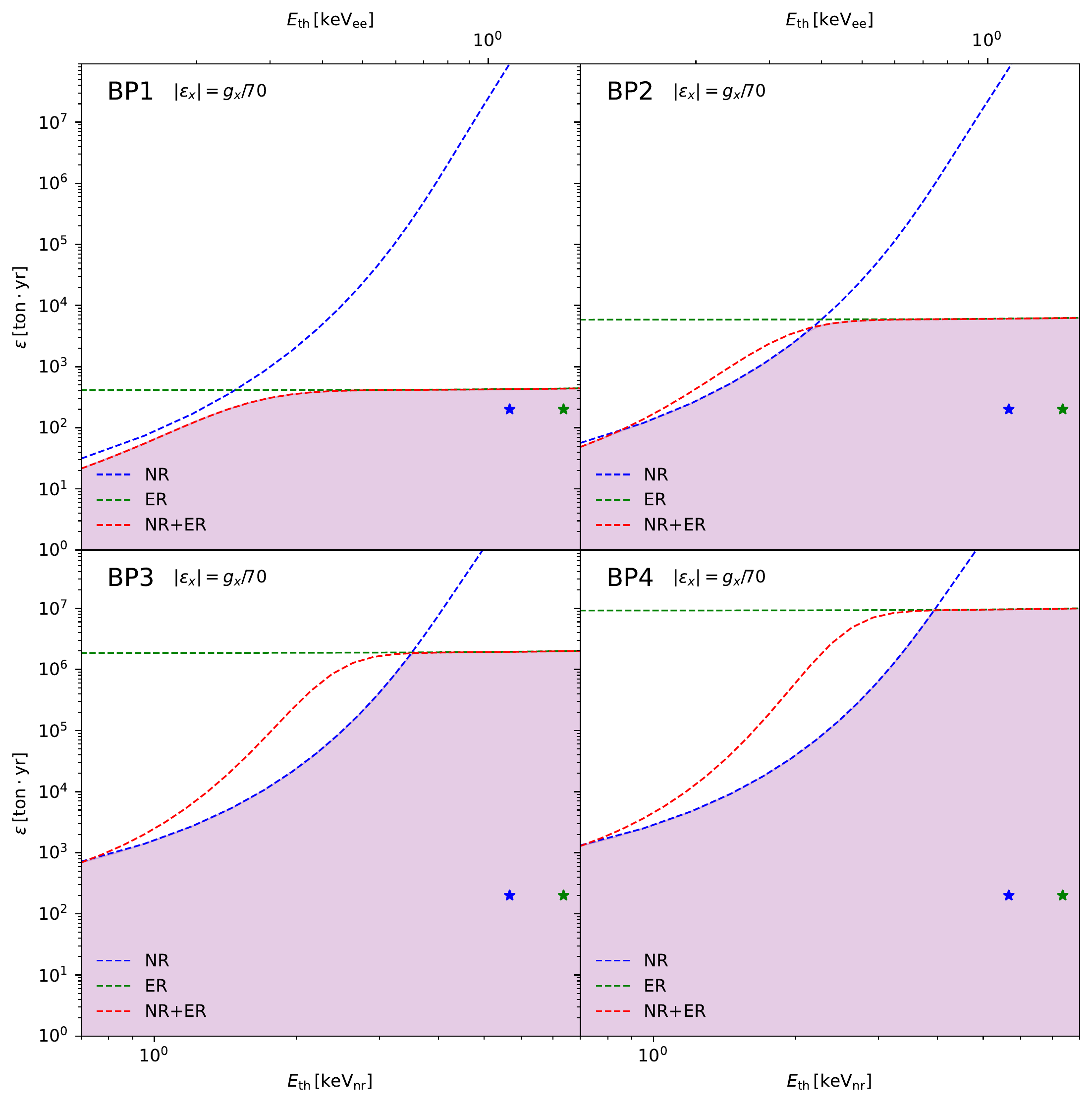}
    \end{center}
    \caption{DARWIN discovery region for searches based on NR (blue dashed line) ER (green dashed line) and NR+ER (red dashed line) as a function of the exposure ($\epsilon$) and energy threshold ($E_{\rm th}$) for each benchmark point of \cref{tab:bps}. In the area above each of these lines, the experimental setup would allow a $5\sigma$ evidence with respect to the SM prediction in each type of search. The nominal DARWIN configurations for nuclear recoils and electron recoils are shown by means of a blue and a green asterisk, respectively.}
    \label{fig:expo-th}
\end{figure*}

\begin{figure*}[!t]
    \begin{center}
        \includegraphics[width=0.9\textwidth]{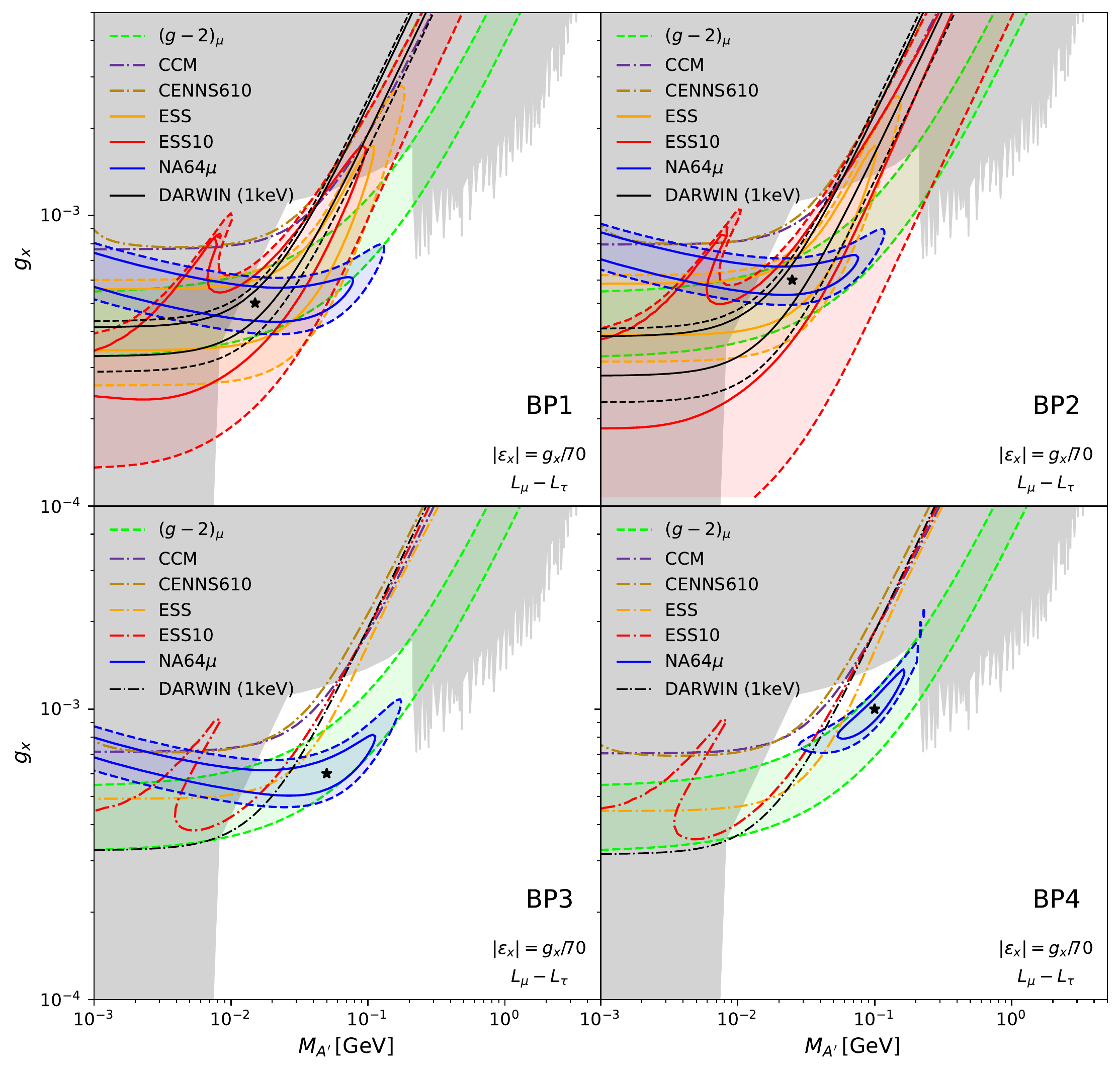}
    \end{center}
    \caption{Parameter reconstruction for each of the BPs of  \cref{tab:bps} assuming a \Umt model. We use direct detection data from a hypothetical low-threshold search ($E_{\rm th}=1$~keV) in the future DARWIN detector. For BP1, the black solid (dashed) contours correspond to the 68\% CL (95\% CL) contours. For the rest of the BPs, the black dot-dashed lines represent the 90\% CL upper limit on the coupling $g_x$. The rest of the lines follow the same convention of \cref{fig:cevns+na64_mutau}. The black star marks the benchmark point. }
    \label{fig:na64+coherent+dd}
\end{figure*}

We have only shown the predictions for BP1, as that is the most optimistic case. The differential rates for BP2 are a bit smaller and for BP3 and BP4 they lie very close to the SM, being much more difficult to separate.

In order to determine the potential of direct detection experiments to disentangle \Umt and \Um, we have computed the number of expected events using the experimental configurations of future liquid xenon detectors for the \Umt model in the four benchmark points of \cref{tab:bps}. We find that for the upcoming LZ and XENONnT, the number of counts in both the NR and ER channels will be too small to be able to probe any of the benchmark points at the $5\,\sigma$ level.

We have then considered the exposure and energy threshold as free parameters and studied the minimal conditions under which a $5\sigma$ observation of \Umt could be claimed at each benchmark point. To compute the significance, we construct the (one-tail) $p$-value,
\begin{equation}
    p \equiv \sum_{n \geq n_\mathrm{obs}} \frac{b^n e^{-b}}{n!}\,,
\end{equation}
where $b$ is the number of background events, and $n_{\rm obs}$ is the observed number of events (in our case the theoretical expectation for each BP). A one-tail statistic is used as we expect the number of observed events to be only in excess of the background-only result. For a discovery-level significance of $5\sigma$, we require a $p$-value of $2.87\times 10 ^{-7}$.

In our analysis, we solve for the required threshold-exposure pairs needed to produce a discovery-level measurement of \Umt for each of our benchmark points. We take the background-only hypothesis to contain all of the counts from SM \cevns and experimental background events. We repeat the analysis individually for each type of search: NR, ER and NR+ER. When lowering the experimental threshold, we assume each experiment's respective efficiencies would improve in a way akin to an extension of their original efficiency functions in log-space, such that their new, lower thresholds would be where the now extended efficiency function reaches 50\%. In all cases, we take the maximum of the energy window to reside at $30\,\kee$ ($\sim 13\,\knr$), above which the double-$\beta$ decay of $^{136}$Xe is expected to dominate over the solar neutrino signal \cite{Mount:2017qzi}.

When lowering the thresholds, we have been careful not to allow the experimental resolutions to increase to arbitrarily large values at low energies, making a reconstruction of low-energy thresholds unrealistic. We have therefore capped each resolution function at its value at the original 50\% threshold, shifting this capped resolution linearly when lowering the thresholds. To ensure a valid conversion of the resolution functions into NR energy scales, we have taken $0.7\,\knr$ ($\sim 0.1 \kee$) as a minimum threshold, corresponding to the lowest energy at which the Lindhard model has been experimentally verified for LXe detectors \cite{Akerib:2016mzi}.

\cref{fig:expo-th} shows the resulting values of the exposure and threshold energy for which a $5\sigma$ discovery of \Umt could be claimed. The area above the dashed blue line would be accessible via NR, the area above the green dashed line corresponds to ER and the region above the red line can be observed via NR+ER. This figure illustrates that the optimal strategy to observe NR is to decrease the threshold, whereas to detect ER one simply needs a bigger target. In the case of BP1 (the only benchmark point for which there is a real chance of observation), a 100~ton~yr exposure with a threshold of 1 keV would suffice. This exposure is comparable to that of the projected DARWIN observatory. Regarding the threshold, it should be pointed out that LUX has demonstrated nuclear recoil calibration for energies down to 1~keV while allowing for NR/ER discrimination  \cite{Akerib:2016mzi}, so we consider this to be a realistic configuration. This figure suggests that one could also aim for a larger threshold in order to explore new physics using ER (for BP1 one would require 400~ton~yr). However, as we explained above, both \Umt and \Um predict a similar increase in the number of events and this search would not allow us to disentangle both models. The discovery lines for LZ and XENONnT do not seem achievable in their experimental configurations; we have included them in~\cref{sec:disco} for completeness.

Finally, using  this hypothetical low-threshold ($E_{\rm th}=1$~keV) configuration for the future DARWIN detector, we have explored how future direct detection results can help reconstruct the parameters of the \Umt model. \cref{fig:na64+coherent+dd} shows the reconstruction of parameters in the $(g_x, M_{A'})$ plane when direct detection data is added to the results from searches at \NAmu and spallation sources, when these are interpreted in terms of a \Umt model. In line with our analysis of \cref{fig:expo-th}, the $5\,\sigma$ discovery of BP1 or BP2, when combined with data from \NAmu and spallation sources, would considerably narrow down the area compatible with \gmu, reducing the uncertainty on both the coupling and the dark photon mass. Notice that, if we were to interpret the direct detection results in terms of a \Um scenario, the whole parameter space for BP1 and BP2 would be ruled out\footnote{Only an extremely fine-tuned region of the parameter space survives where, due to the interference term, the \Um contribution turns positive. This only happens for large values of the coupling ($g_{\mu}\sim 10^{-3}$) and is therefore well within the area of the parameter space ruled out by existing experimental bounds. }.

For BP3 and BP4, direct detection will not have enough sensitivity to claim detection. Nevertheless, the bounds derived from direct searches, would further constrain the parameter space. Similar to \cevns bounds, direct detection will be more effective in constraining mediator masses below approximately 50~MeV in the \gmu region. These bounds will be similar for both \Umt and \Um scenarios and no extra information would be gained in that respect.

\section{Conclusions}
\label{sec:conclusions}

In this article, we have explored the complementarity of different experimental probes of muon-philic solutions to the $4.2\,\sigma$ deviation between the observed value of the muon anomalous magnetic moment at the E989 experiment and the SM theoretical prediction. In particular, we have focused on the light vector mediator that arises from the anomaly-free \Umt model. We have laid out a strategy of how to combine muon beam experiments, \cevns experiments at spallation sources and DM direct detection experiments to confirm whether the observed \gmu excess is indeed due to a \Umt boson. We have done so by contrasting our findings for \Umt to those of a phenomenological \Um model, which can experimentally mimic many but not all of the properties of an \Umt. 

We can summarise our findings as follows:

\begin{itemize}
    \item[$\bullet$] 
    \NAmu will be the first experiment to be sensitive to the \gmu region in \Um and \Umt. For mediator masses of the order of $M_{A'}\sim100$~MeV, it would allow for an excellent reconstruction of both the coupling $g_x$ and the mass $M_{A'}$. For lighter hidden photons ($M_{A'}\lesssim 50$~MeV) it can still give a good reconstruction of the coupling while only providing an upper bound on the mediator mass. In general, it is, however, insensitive to the kinetic mixing $\epsilon_x$. Furthermore, with data from \NAmu alone the models \Umt and \Um cannot be discriminated (nor can they be distinguished from a muon-coupled mediator that can decay into light DM states). \\
    
    \item[$\bullet$]
    Future experiments looking for \cevns at spallation sources, such as the planned CENNS610 and CCM, will set constraints on the low-mediator mass ($M_{A'}\lesssim50$~MeV) region of the \gmu solution.
    The predicted \cevns rate for the \Umt and \Um models is smaller than in the SM, making this observation more challenging. 
    Large detectors near powerful sources such as the projected experiment at the ESS or smaller devices with an extremely low-threshold (we consider a $E_{\rm th}=0.1$~keV version of ESS with just 10~kg) could be able to reconstruct the coupling and set upper bounds on the mediator mass.
 
    However, since these experiments only test the coupling of the hidden photon with the muon sector through the combination $g_x\epsilon_x$, they will be unable to discriminate the \Umt and \Um models on their own. \\

    \item[$\bullet$]
    From the combination of data from \NAmu (which measures $g_x$) and \cevns experiments (which measure $g_x\epsilon_x$), one can infer the value of the kinetic mixing and also improve the reconstruction of the mediator mass. However, this would be insufficient to distinguish the \Umt and \Um model as, in principle, the kinetic mixing could be the same ($\epsilon_x=-g_x/70$). \\
    
    \item[$\bullet$]
    Direct detection experiments will provide a unique opportunity to test the couplings to the tau sector via scattering off solar $\nu_\tau$'s. 
    In particular, because of the contribution of tau neutrinos, the expected nuclear recoil rate from \cevns in \Umt is larger than the SM prediction, whereas for the \Um with $\epsilon_\mu=-g_{\mu}/70$ a decrease is expected.
    This allows for discrimination of both solutions to \gmu in the values of the kinetic mixing to which \NAmu and spallation experiments are most insensitive.
    We have estimated the experimental requirements to probe the low-mass window of the \gmu region. We conclude that a liquid xenon experiment with a total exposure of 100~ton~yr and a threshold of $E_{\rm th}=1$~keV would suffice to probe mediator masses below $25$~MeV. These are similar characteristics to those of the planned DARWIN detector. The mass and coupling reconstruction in that area would be complementary to that of \NAmu and \cevns experiments.
    
    In the event of no observation, DARWIN (low threshold) would be able to rule out the low-mass window of the \gmu parameter space.

\end{itemize}

Our results show very promising prospects for separating a potential \gmu signal of a \Umt from a \Um boson through the combination of future data from \NAmu, \cevns experiments at spallation sources and, crucially, direct dark matter detectors. This strategy would allow us to confirm \Umt as the solution to the \gmu excess.

\section*{Acknowledgements}

We thank Felix Kahlhoefer for helpful discussions on the resolution of direct detection experiments. We further want to thank Yue Zhang for pointing out stringent flavour constraints on \Um bosons. DGC acknowledges financial support from the project SI2/PBG/2020-00005 and is also supported in part by the Spanish Agencia Estatal de Investigaci\'on through the grants PGC2018-095161-B-I00 and IFT Centro de Excelencia Severo Ochoa SEV-2016-0597, and the Spanish Consolider MultiDark FPA2017-90566-REDC. PF is funded by the UK Science and Technology Facilities Council (STFC) under grant ST/P001246/1. AC is supported by the F.R.S.-FNRS under the Excellence of Science EOS be.h project n. 30820817. Computational resources have been provided by the supercomputing facilities of the Université Catholique de Louvain (CISM/UCL) and the Consortium des Équipements de Calcul Intensif en Fédération Wallonie Bruxelles (CÉCI) funded by the Fond de la Recherche Scientifique de Belgique (F.R.S.-FNRS) under convention 2.5020.11 and by the Walloon Region.
\clearpage

\appendix
\onecolumn
\section{Discovery Lines for LZ and XENONnT}
\label{sec:disco}

\begin{figure*}[h]
    \begin{center}
    \includegraphics[width=0.9\textwidth]{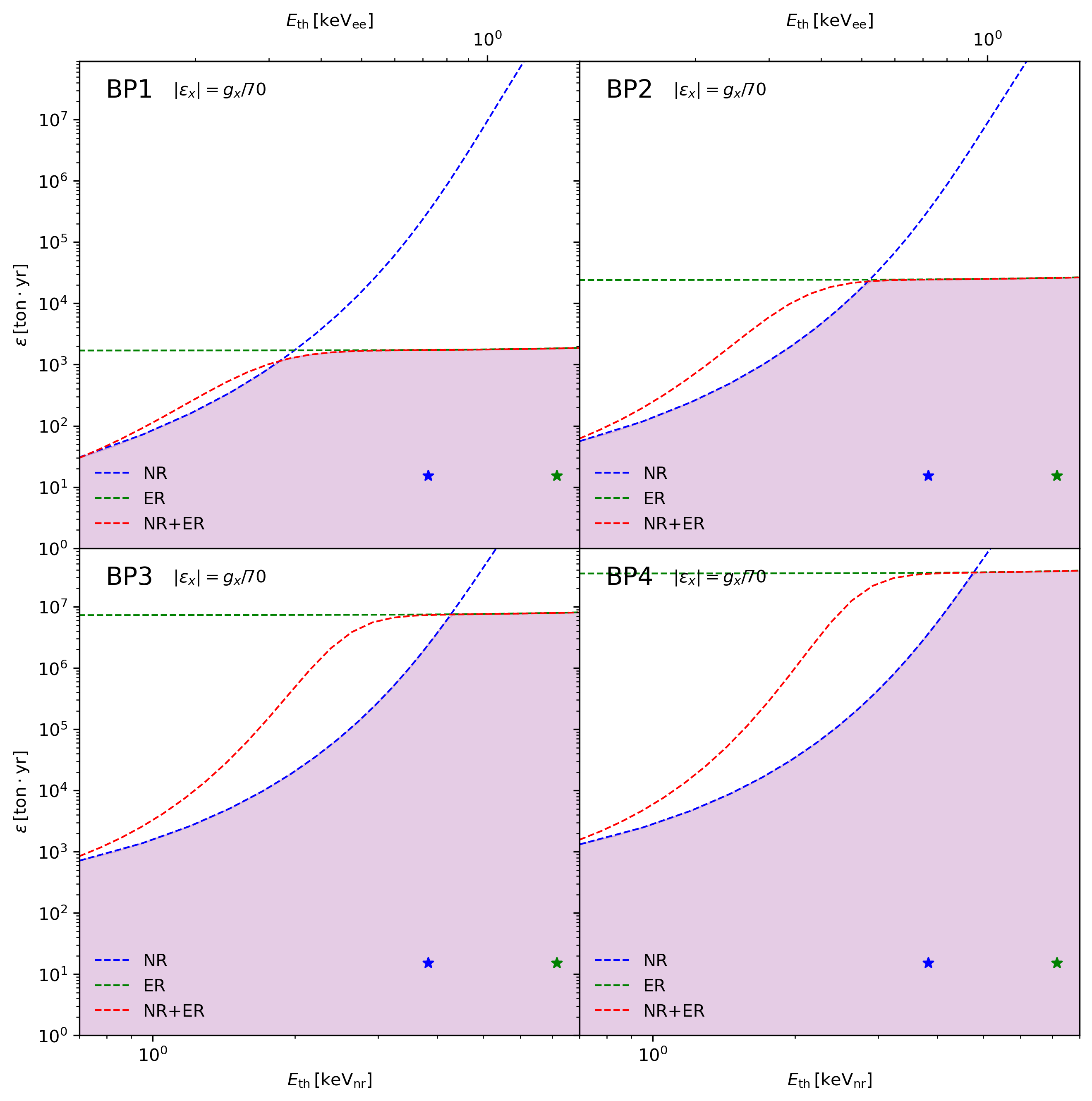}
    \end{center}
    \caption{The same as in \cref{fig:expo-th} but for LZ.
    }
    \label{fig:expo-th_lz}
\end{figure*}

\begin{figure*}[!t]
    \begin{center}
    \includegraphics[width=0.9\textwidth]{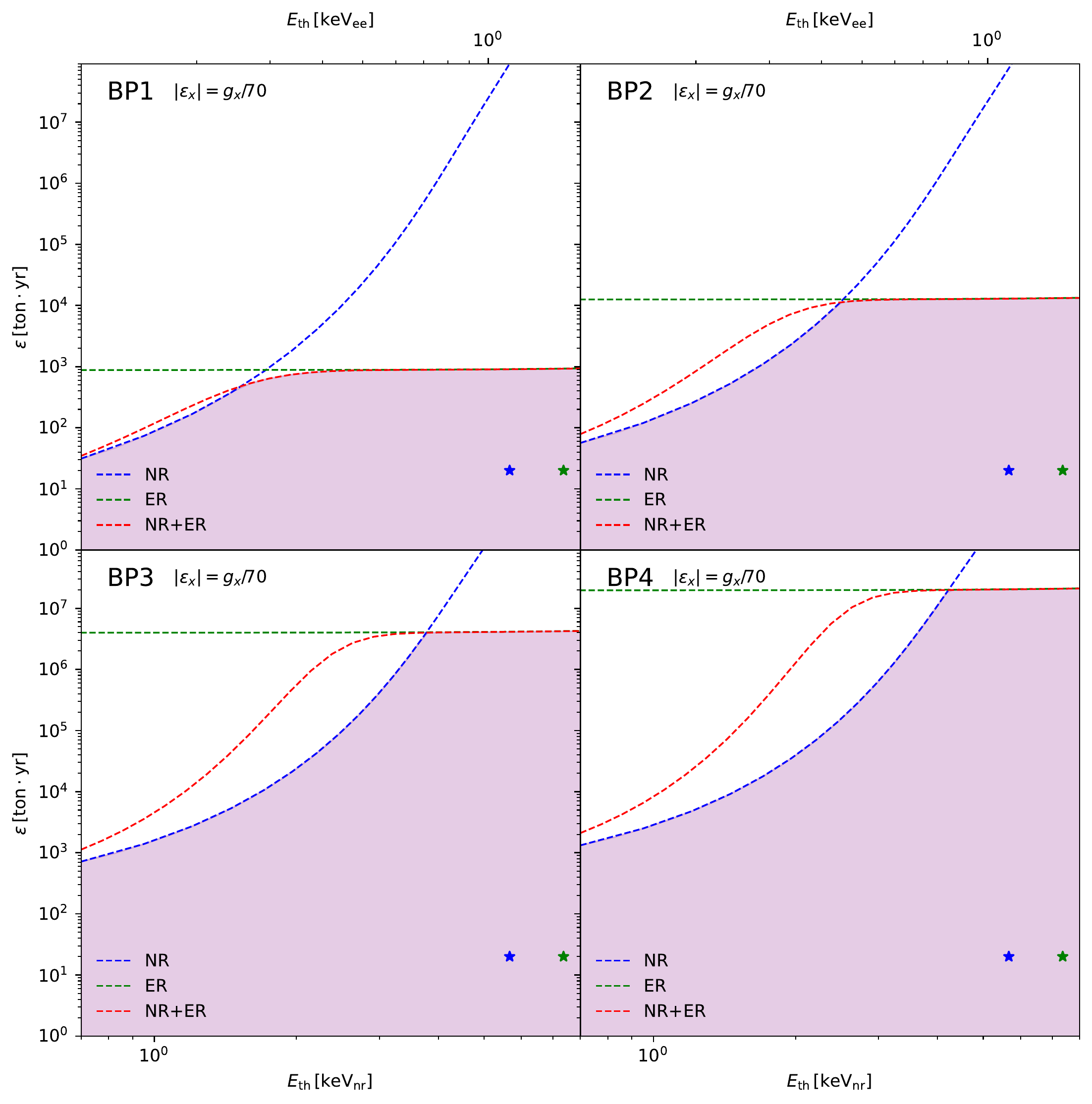}
    \end{center}
    \caption{The same as in \cref{fig:expo-th} but for XENONnT. 
    }
    \label{fig:expo-th_xnt}
\end{figure*}

In this appendix, we show the 5$\sigma$ discovery lines for LZ and XENONnT in Fig.~\ref{fig:expo-th_lz} and Fig.~\ref{fig:expo-th_xnt}, respectively. Please refer to Sec.~\ref{sec:direct} for details.

\twocolumn

\bibliographystyle{JHEP}
\bibliography{literature}

\end{document}